\documentclass[aip,jcp,reprint,groupedaddress]{revtex4-1} 
\usepackage[english]{babel}

\usepackage{amsmath, amssymb}
\usepackage{mathtools}
\usepackage{siunitx}
\usepackage{environ}
\usepackage{braket}
\usepackage{bm}
\DeclareMathAlphabet{\mathscrlower}{OT1}{pzc}{m}{it} 
\usepackage{prettyref}
\usepackage{booktabs}
\usepackage{multirow}
\usepackage{threeparttable}
\usepackage{threeparttablex}

\newcommand{\pauli}{\boldsymbol{\sigma}}
\newcommand{\Pauli}{\boldsymbol{\sigma}}
\newcommand{\diraccontra}[1]{\boldsymbol{\gamma}^{#1}}

\newcommand{\diraca}{\vec{\boldsymbol{\alpha}}}

\newcommand{\unity}{\bm{1}_{2\times 2}}
\newcommand{\zeroty}{\bm{0}_{2\times 2}}
\newcommand{\pos}{\vec{r}}
\newcommand{\mom}{\vec{p}}
\newcommand{\momop}{\hat{\vec{p}}}
\newcommand{\momoppot}{\hat{\vec{\pi}}}
\newcommand{\spinmom}{\vec{\Pauli}\cdot\momop}
\newcommand{\spinmompot}{\vec{\Pauli}\cdot\momoppot}
\newcommand{\efield}{\mathcal{E}}
\newcommand{\bfield}{\mathcal{B}}
\newcommand{\Sum}[2]{\sum\limits_{#1}^{#2}}
\newcommand{\parantheses}[1]{\left(#1\right)}
\newcommand{\brackets}[1]{\left[#1\right]}
\newcommand{\braces}[1]{\left\{ #1\right\}}
\let\nablatmp\nabla
\renewcommand{\nabla}{\vec{\nablatmp}}
\DeclarePairedDelimiter\abs{\lvert}{\rvert}
\makeatletter
\let\oldabs\abs
\def\abs{\@ifstar{\oldabs}{\oldabs*}}

\newcommand{\Op}[1]{\hat{#1}}

\newcommand{\partiell}[2]{\frac{\partial #1}{\partial #2}}

\AtBeginDocument{
\heavyrulewidth=.08em
\lightrulewidth=.05em
\cmidrulewidth=.03em
\belowrulesep=.65ex
\belowbottomsep=0pt
\aboverulesep=.4ex
\abovetopsep=0pt
\cmidrulesep=\doublerulesep
\cmidrulekern=.5em
\defaultaddspace=.5em
}
\begin{document}
\title{Zeroth order regular approximation approach to electric dipole
moment interactions of the electron}
\date{\today}
\author{Konstantin Gaul}
\author{Robert Berger}
\affiliation{Fachbereich Chemie, Philipps-Universit\"{a}t Marburg,
Hans-Meerwein-Stra\ss{}e 4, 35032 Marburg, Germany}
\begin{abstract}
A quasi-relativistic two-component approach for an efficient calculation of
$\mathcal{P,T}$-odd interactions caused by a permanent electric dipole
moment of the electron (eEDM) is presented. The approach uses
a (two-component) complex generalized Hartree-Fock (cGHF) and a complex
generalized Kohn-Sham (cGKS) scheme within the zeroth order regular
approximation (ZORA). In applications to select heavy-elemental polar
diatomic molecular radicals, which are promising candidates for an eEDM
experiment, the method is compared to relativistic four-component
electron-correlation calculations and confirms values for the effective
electrical field acting on the unpaired electron for RaF, BaF, YbF and HgF.
The calculations show that purely relativistic effects, involving only 
the lower component of the Dirac bi-spinor, are well described by treating
only the upper component explicitly.
\end{abstract}
\maketitle
\section{Introduction}
Violations of fundamental symmetries, such as those related to a combined charge conjugation
($\mathcal{C}$) and parity ($\mathcal{P}$) operation
($\mathcal{CP}$-violation), provide stringent tests for physics beyond the
Standard Model of particle physics, which is often referred to as new
physics.\cite{gross:1996,fortson:2003} The permanent electric dipole moment
(EDM) of particles, which is the target of the present work, emerges from
violation of both parity ($\mathcal{P}$) and time-reversal ($\mathcal{T}$)
symmetry that is related to $\mathcal{CP}$-violation via the
$\mathcal{CPT}$-theorem\cite{schwinger:1951,pauli:1955,luders:1954}.\cite{khriplovich:1997}\par
 
Even before the first experimental evidence of $\mathcal{CP}$-violation in
Kaon decays,\cite{christenson:1964} Salpeter studied the effect of a
permanent EDM of an electron (eEDM) and presented first calculations for
hydrogen-like atoms in 1958.\cite{salpeter:1958} Sanders found strong
relativistic enhancement of effects due to an eEDM in the late
1960s\cite{sandars:1965,sandars:1966,sandars:1968,sandars:1968a} and
suggested the use of polar diatomic molecules that contain a high $Z$
element (with $Z$ being the nuclear charge) for the search of the proton
EDM.\cite{sandars:1967} In the 1970s Labzowsky,\cite{labzowsky:1978}
Gorshkov \textit{et. al}\cite{gorshkov:1979} and Sushkov and
Flambaum\cite{sushkov:1978} showed that effects become large in polar heavy
diatomic radicals. Some years later Sushkov, Flambaum and
Khriplovich\cite{sushkov:1984,flambaum:1985} suggested to exploit
small $\Omega$-doubling effects in $^2\Sigma_{1/2}$-ground states of
molecules such as HgF or BaF, which were first studied theoretically
by Kozlov in a
semi-empirical model in 1985.\cite{kozlov:1985} Recently, in 2014, the
currently lowest upper limit on the eEDM ($d_\text{e}\lesssim8.7\times10^{-29}~e\cdot$~cm)
was measured using the $H^3\Delta_1$-state of ThO.\cite{baron:2014} As
$\mathcal{CP}$-violation in the Standard Model is only embedded on the
level of quarks and gives rise to an EDM in the lepton sector only via
higher order radiative corrections, an eEDM is a sensitive probe for
new physics.\cite{bernreuther:1991,nir:2000}\par

The theoretical search for still more favourable candidates for eEDM
experiments, employing relativistic quantum chemistry, is
vivid.\cite{kudashov:2014,skripnikov:2015,prasannaa:2015,denis:2015,fleig:2016,sasmal:2016,sasmal:2016a,sunaga:2016}
So far, besides the strong $Z$-dependence of $\mathcal{CP}$-violating
effects, there appears to be no thorough understanding of the mechanisms
that make eEDM enhancement in molecular systems large, but first attempts
in this direction have been reported recently\cite{sunaga:2017}. Thus, a
systematic study of $\mathcal{P,T}$-odd effects in molecules is of avail.
Yet relativistic many-electron calculations are computationally demanding.
The main effort in relativistic calculations stems from an explicit
consideration of the small component of the Dirac wave
function.\cite{dyall:2007} Quasi-relativistic approaches, such as the
zeroth order regular approximation (ZORA), improve on computation time
considerably and perform very well in most molecular
calculations.\cite{leeuwen:1994} Even in the description of relativistic
molecular properties such as parity violating effects in molecules ZORA
proved to be very reliable.\cite{berger:2005,berger:2005a,berger:2007,
berger:2008,nahrwold:09,isaev:2012,isaev:2013,isaev:2014}

\section{Methodology}
\subsection{Theory of eEDM Interactions in Molecules} 
Salpeter introduced a perturbation of the Dirac equation due to
interactions with an electric dipole moment of an electron to describe 
permanent EDMs of atoms.\cite{salpeter:1958} 
The Lorentz- and gauge-invariant formulation appears as
\begin{multline}
\brackets{\diraccontra{\mu}\parantheses{\i\hbar\partial_\mu + \frac{e}{c}A_\mu}-m_\text{e}c{\bf 1}_{4\times4}}\psi\\
=-\frac{d_\text{e}}{2}\diraccontra{5}\underline{\underline{\bm{\sigma}}}^{\mu\nu}F_{\mu\nu}\psi .
\label{eq: loiveedm}
\end{multline} 
Here $c$ is the speed of light in vacuum, $e$ is the electric
constant, $m_\text{e}$ is the mass of the electron, $\i=\sqrt{-1}$ is the
imaginary unit, $d_\text{e}$  is the eEDM, $\psi$ is a Dirac four-spinor, 
$A_\mu=\parantheses{\Phi,-\vec{A}}$ is the four-potential with the
scalar and vector potentials $\Phi$ and $\vec{A}$,
$\partial_\mu=\partiell{}{\mathrm{x}_\mu}$ is a first derivative of
the four vector $\mathrm{x}_\mu=(ct,x,y,z)$, 
$\diraccontra{\mu}$, $\diraccontra{5}$ are the Dirac matrices in
standard notation:
\begin{equation}
\begin{gathered}
\diraccontra{0}=\begin{pmatrix}
\unity & \zeroty\\
\zeroty & -\unity 
\end{pmatrix},~~
\diraccontra{k}=\begin{pmatrix}
\zeroty&\pauli^k\\
-\pauli^k&\zeroty
\end{pmatrix}\\
\diraccontra{5}=\i\diraccontra{0}\diraccontra{1}\diraccontra{2}\diraccontra{3}
\end{gathered}
\end{equation}
and
$\underline{\underline{\bm{\sigma}}}^{\mu\nu}=\brackets{\diraccontra{\mu},\diraccontra{\nu}}_-/2$,
with $\brackets{a,b}_-=ab-ba$ being the commutator.
$F_{\mu\nu}=\partial_\mu A_\nu-\partial_\nu A_\mu$ is
the field strength tensor.
In the above equations index notation ($\mu,\nu=0,1,2,3;~k=1,2,3$) and Einstein's sum convention are employed.
 After evaluation of the  tensor product  $\underline{\underline{\bm{\sigma}}}^{\mu\nu}F_{\mu\nu}$ the left-hand side of Eq. \prettyref{eq: loiveedm} reduces to a term proportional to the four-component analogue of the Pauli spin-matrix vector $\vec{\boldsymbol{\Sigma}}=\unity\otimes\vec{\pauli}$, representing the interaction with the electric field $\vec{\efield}$ and a term proportional to $\i\diraca$, where $\bm{\alpha}^{k}=\diraccontra{0}\diraccontra{k}$ with $k=1,2,3$, representing the interaction with the magnetic field $\vec{{B}}$. 
Thus the Lorentz invariant eEDM Hamiltonian has the form
\begin{equation}
\Op{H}_{\text{eEDM}}=-d_\text{e}\diraccontra{0}\brackets{\vec{\boldsymbol{\Sigma}}\cdot\vec{\efield}
+\i\diraca\cdot\vec{\bfield}}.
\end{equation} 
It was shown that the magnetic term gives minor contributions in many-body calculations and in leading order this Hamiltonian further reduces to the interaction with the electric field:\cite{lindroth:1989}
\begin{equation}
\Op{H}_{\text{eEDM}}\approx-d_\text{e}\diraccontra{0}\vec{\boldsymbol{\Sigma}}\cdot\vec{\efield}.
\label{eq: eedmop}
\end{equation} 
In 1963 Schiff stated, that in the non-relativistic limit the expectation value of  $\Op{H}_{\text{eEDM}}$ of an atom vanishes, independent whether the elementary particles in the atom have an EDM or not.\cite{schiff:1963} Therefore an atom in the non-relativistic limit has always a zero EDM. Addition of $0=d_\text{e}\parantheses{\vec{\boldsymbol{\Sigma}}\cdot\vec{\efield}(\vec{r})
-\vec{\boldsymbol{\Sigma}}\cdot\vec{\efield}(\vec{r})}$ results in the alternative formulation\cite{sandars:1965,sandars:1966,sandars:1968,sandars:1968a}
\begin{equation}
\Op{H}_{\text{eEDM}}=-d_\text{e}\vec{\boldsymbol{\Sigma}}\cdot\vec{\efield}(\vec{r})-d_\text{e}\parantheses{\diraccontra{0}-1}\vec{\boldsymbol{\Sigma}}\cdot\vec{\efield}(\vec{r}),
\end{equation} 
with the total electrical field $\vec{\efield}=\vec{\efield}_\text{int}+\vec{\efield}_\text{ext}$ being a sum of the internal and external fields $\vec{\efield}_\text{int}$, $\vec{\efield}_\text{ext}$. 
Whereas the first term on the right is zero due to Schiff's theorem, the second appears only in the small component of the Dirac equation, because
\begin{equation}
\parantheses{\diraccontra{0}-1}=-2\cdot\begin{pmatrix}
\zeroty&\zeroty\\
\zeroty&\unity
\end{pmatrix}.
\end{equation}
Since the momentum operator commutes with the Dirac-Hamiltonian,  the effective Hamiltonian for the eEDM can be reformulated by commuting the unperturbed Hamiltonian with a modified momentum operator and defining an effective operator $\Op{H}_\text{d}$:
\begin{eqnarray}
&\Op{H}_\text{tot}=\Op{H}+\Op{H}_{\text{eEDM}}\\
&\Op{H}_{\text{eEDM}}=\brackets{\hat{P},\Op{H}}+\Op{H}_\text{d}.
\end{eqnarray}
In consistency with the notation of Lindroth, Lynn and Sandars\cite{lindroth:1989} the modified momentum operator of the above derivation is called $\hat{P}_\text{I}$  and the reformulation of $\Op{H}_{\text{eEDM}}$ using $\hat{P}_\text{I}$ is denoted as Stratagem I, which can be summarized as
 \begin{eqnarray}
&\qquad\hat{P}_\text{I}\equiv-\frac{\i d_\text{e}}{\hbar e}\Sum{i=1}{N_\text{elec}}\vec{\boldsymbol{\Sigma}}\cdot\momop_i \\
&\Rightarrow~\Op{H}_{\text{d},\text{I}}=-d_\text{e}\Sum{i=1}{N_\text{elec}}\parantheses{\diraccontra{0}-1}\vec{\boldsymbol{\Sigma}}\cdot\vec{\efield}(\vec{r}_i),
\label{eq: stratagemI}
\end{eqnarray}
Here $\momop$ is the linear momentum operator, $\hbar=h/(2\pi)$ is the reduced Planck constant and the sum runs over all $N_\text{elec}$ electrons of system.
Additionally, introducing a factor of $\diraccontra{0}$ in the modified momentum, an alternative expression, called Stratagem II, can be derived:\cite{martensson-pendrill:1987,commins:2010}
 \begin{eqnarray}
&\qquad\hat{P}_\text{II}\equiv-\frac{\i d_\text{e}}{\hbar e}\Sum{i=1}{N_\text{elec}}\diraccontra{0}\vec{\boldsymbol{\Sigma}}\cdot\momop_i\\
&\Rightarrow~\Op{H}_{\text{d},\text{II}}=\frac{2\i c d_\text{e}}{\hbar e}\Sum{i=1}{N_\text{elec}} \diraccontra{0}\diraccontra{5}\momop_i^2.
\label{eq: stratagemII}
\end{eqnarray}
This operator has, within the Dirac-Coulomb picture, the advantage of
being a single-particle operator, whereas
$\Op{H}_{\text{d},\text{I}}$ is due to the internal electrical field, which is a
function of the Coulomb potential $V_\text{C}$, a many-body operator
(see below). Lindroth, Lynn and Sandars have pointed out, that effects of the Breit
interaction and one-photon exchange, which are on the same order,
namely $\mathcal{O}(\alpha^2)$, give only minor contributions of less
than one percent.\cite{lindroth:1989}\par
For the evaluation of $\Op{H}_{\text{d},\text{I}}$ the mathematical form of the electrical field is of interest. 
Starting from its definition 
\begin{equation}
\vec{\efield}_\text{int}(\pos_i)\equiv\frac{1}{e}\nabla V_\text{C}(\pos_i),
\end{equation}
using the molecular or atomic Coulomb potential and assuming a spherically symmetric charge distribution of the nucleus, e.g. a Gaussian charge distribution, the Gau\ss\ law is valid and the internal electrical field is
 \begin{equation}
\vec{\efield}(\pos_i)=\Sum{A=1}{N_\text{nuc}}\frac{Z_A
e}{4\pi\epsilon_0}\frac{\pos_i-\pos_A}{\abs{\pos_i-\pos_A}^3}-\Sum{j\not=i}{N_\text{elec}}\frac{e}{4\pi\epsilon_0}\frac{\pos_i-\pos_j}{\abs{\pos_i-\pos_j}^3},
\end{equation}
with the vacuum permittivity $\epsilon_0$, the number of protons $Z_A$
of nucleus $A$, the vector in position space $\pos$ and the sums
running over all $N_\text{nuc}$ nuclei of the molecule. 
The first term on the right-hand side arises from the electrostatic
fields of the nuclei, experienced by the electron, and the second term
arises from the electrostatic fields of the other electrons. Due to
the latter, $\Op{H}_{\text{d},\text{I}}$ is a many-body Hamiltonian and
therefore much more difficult to treat in numerical calculation.\par
Fortunately, it has been shown, that the two-electron contribution is
on the order of one percent only in relativistic many-body
calculations and thus it is typically well justified to drop this
term\cite{lindroth:1989}. 
The internal electrical field reduces then to the nuclear contribution
only:
 \begin{equation}
\vec{\efield}(\pos_i)\approx\Sum{A=1}{N_\text{nuc}}\frac{Z_A e}{4\pi\epsilon_0}\frac{\pos_i-\pos_A}{\abs{\pos_i-\pos_A}^3}.
\label{eq: efieldapprox}
\end{equation}
In this approximation $\Op{H}_{\text{d},\text{I}}$ is a single-particle
operator and for this reason there is no longer an advantage in using
$\Op{H}_{\text{d},\text{II}}$ instead, even within the Dirac-Coulomb picture.

In this work both forms will be used to derive a quasi-relativistic
theory in the zeroth order regular approximation (ZORA) framework.
This does not only allow a comparison of the two transformations of
the perturbed Dirac equation, but also provides a test of the newly
developed quasi-relativistic approach, as both stratagems should
yield approximately the same results. 

\subsection{Derivation of the eEDM ZORA equations}
In the following we will derive the ZORA eEDM interaction Hamiltonian, starting from the Stratagem I Hamiltonian $\Op{H}_{\text{d},\text{I}}$ as perturbation to the molecular Dirac equation  (see Equation \prettyref{eq: stratagemI}) and repeat the derivation afterwards for $\Op{H}_{\text{d},\text{II}}$, received from Stratagem II (see Equation \prettyref{eq: stratagemII}).
\subsubsection{Derivation starting from Stratagem I}
The $\Op{H}_{\text{d},\text{I}}$ perturbed molecular Dirac equation can be written in block matrix form as
\begin{widetext}
\begin{equation}
\begin{pmatrix}
\Op{V}(\pos_i)-\epsilon&c\spinmompot_i\\
c\spinmompot_i &\Op{V}(\pos_i)-\epsilon-2m_\text{e}c^2+2d_\text{e}\vec{\Pauli}\cdot\vec{\efield}(\pos_i)\\
\end{pmatrix}
\begin{pmatrix}
\phi(\pos_i)\\
\chi(\pos_i)\\
\end{pmatrix}
=
\begin{pmatrix}
\vec{0}\\
\vec{0}\\
\end{pmatrix},
\end{equation}
\end{widetext}
where $\phi$, $\chi$ are the large and small components of the Dirac
spinor with energy $\epsilon$ shifted by $m_\text{e}c^2$,
$\momoppot=\momop+e\vec{A}$ is the minimal coupling relation
for the electron with $\vec{A}$ being the vector potential, $\Op{V}$
is the scalar potential energy operator and  $i$ is the index of the
electron, which will be dropped in the following for better
readability.\par
Now using the elimination of small component (ESC) method,\cite{chang:1986} the small component can be expressed as
\begin{multline}
\chi(\pos)={\underbrace{\parantheses{\underbrace{\parantheses{2m_\text{e}c^2-\Op{V}(\pos)+\epsilon}\unity}_{\bf A}\underbrace{-2d_\text{e}\vec{\Pauli}\cdot\vec{\efield}(\pos)}_{\text{d}_\text{e}{\bf B}}}}_{\bf M}}^{\;-1}\\
\times c\parantheses{\spinmompot}\phi(\pos).
\end{multline}
This equation is valid only if the matrix $\bf M$ is invertible. This is
commonly considered to be the case, because $\bf B$ has the form of the Pauli
matrices (which are invertible) and $\bf A$ is a diagonal matrix and
therefore  does not change the invertibility of $\bf B$ by addition
provided that the parameters of the fields involved are assumed to be
confined appropriately.\par
The inversion of the matrix $\bf M$ can now be rewritten in several ways. The result of the inversion should have the following properties: (i) it should be divisible into an unperturbed and a perturbed Hamiltonian, (ii) it should be expandable in $d_\text{e}$ and (iii) the perturbing leading order term should be linear in $d_\text{e}$.\par
This can be achieved by first extracting ${\bf A}^{-1}=\parantheses{2m_\text{e}c^2-\Op{V}(\vec{r})-\epsilon}^{-1}\unity$, using the matrix relation
\begin{equation}
({\bf A} + d_\text{e}{\bf B })^{-1}={\bf A}^{-1}(\unity+ d_\text{e}{\bf B }{\bf A}^{-1})^{-1}.
\end{equation}
Now the inverse expression can be expanded in $d_\text{e}$ when for typical field situations $\abs{d_\text{e}{\bf B}{\bf A}^{-1}}<\unity$ is assumed:
\begin{equation}
(\unity+ d_\text{e}{\bf B }{\bf A}^{-1})^{-1}=\Sum{m=0}{\infty}\brackets{-d_\text{e}{\bf B }{\bf A}^{-1}}^m ,
\end{equation}
 and we can write down an expression for the ESC Hamiltonian of infinite order in $d_\text{e}$:
\begin{multline}
\Op{H}^\text{ESC}_{\text{tot,I}}=\Op{V}(\pos)\unity+c^2\parantheses{\spinmompot}\parantheses{2m_\text{e}c^2-\Op{V}(\pos)+\epsilon}^{-1}\\
\times\Sum{m=0}{\infty}\brackets{2d_\text{e}\vec{\Pauli}\cdot\vec{\efield}(\pos)\parantheses{2m_\text{e}c^2-\Op{V}(\pos)+\epsilon}^{-1}}^{m}\parantheses{\spinmompot}.
\end{multline} 
This Hamiltonian can be separated into an unperturbed ESC-Hamiltonian $\Op{H}^\text{ESC}_0$ ($m=0$ term and potential energy term) and a perturbation due to the eEDM. The latter can be reduced to the term linear in $d_\text{e}$, as $\abs{d_\text{e}}$ is very small.
\begin{widetext}
\begin{equation}
\Op{H}^\text{ESC}_{\text{tot,I}}=\Op{H}^\text{ESC}_0+\underbrace{c^2\parantheses{\spinmompot}\parantheses{2m_\text{e}c^2-\Op{V}(\pos)+\epsilon}^{-1}2d_\text{e}\vec{\Pauli}\cdot\vec{\efield}(\pos)\parantheses{2m_\text{e}c^2-\Op{V}(\pos)+\epsilon}^{-1}\parantheses{\spinmompot}}_{\Op{H}^\text{ESC}_{\text{d},\text{I}}}.
\end{equation}
\end{widetext}
For regular approximation\cite{chang:1986,lenthe:1993} we extract $\parantheses{2m_\text{e}c^2-\Op{V}(\vec{r})}^{-1}$ from $\parantheses{2m_\text{e}c^2-\Op{V}(\vec{r})-\epsilon}^{-1}$ and expand in the orbital energy $\epsilon$:
\begin{multline}
\parantheses{2m_\text{e}c^2+\epsilon-\Op{V}(\pos)}^{-1}\\
=\frac{1}{2m_\text{e}c^2-\Op{V}(\pos)}\Sum{k=0}{\infty}\brackets{\frac{-\epsilon}{2m_\text{e}c^2-\Op{V}(\pos)}}^k.
\label{eq: regularexp}
\end{multline}
The ZORA eEDM Hamiltonian, linear in $d_\text{e}$, reads
\begin{multline}
\Op{H}^\text{ZORA}_{\text{d},\text{I}} =c^2\parantheses{\spinmompot}\parantheses{2m_\text{e}c^2-\Op{V}(\vec{r})}^{-1}\\
\times2d_\text{e}\vec{\Pauli}\cdot\vec{\efield}(\pos)\parantheses{2m_\text{e}c^2-\Op{V}(\vec{r})}^{-1}\parantheses{\spinmompot}.
\label{eq: ZORAI}
\end{multline}
As $\brackets{\parantheses{2m_\text{e}c^2-\Op{V}(\vec{r})}^{-1},\vec{\Pauli}\cdot\vec{\efield}(\pos)}_-=0$, a modified ZORA factor can be defined as
\begin{equation}
\omega_{\text{d},\text{I}}(\pos)=\frac{2d_\text{e}c^2}{\parantheses{2m_\text{e}c^2-\Op{V}(\vec{r})}^{2}}
\end{equation}
and the final expression for the ZORA eEDM interaction Hamiltonian reads
\begin{equation}
\Op{H}^\text{ZORA}_{\text{d},\text{I}} =\parantheses{\spinmompot}\omega_{\text{d},\text{I}}(\pos)\vec{\Pauli}\cdot\vec{\efield}(\pos)\parantheses{\spinmompot}.
\end{equation}
This operator is approximately a one-electron operator (see Eq. \prettyref{eq: efieldapprox}). 
From here on we drop now terms depending on the vector potential (thus assuming no magnetic interactions). 
Then the matrix elements within a one-electron basis set $\{\varphi_\lambda\}$ are of the form
\begin{equation}
 \begin{aligned}
{H}^\text{ZORA}_{\text{d},\text{I},\lambda\rho}&=\Braket{\varphi_\lambda|\Op{H}^\text{ZORA}_{\text{d},\text{I}}|\varphi_\rho}\\
&=\Braket{\varphi_\lambda|\parantheses{\spinmom}\omega_{\text{d},\text{I}}(\pos)\vec{\Pauli}\cdot\vec{\efield}(\pos)\parantheses{\spinmom}|\varphi_\rho}.
\end{aligned}
\end{equation}
The momentum operators in position space are differential operators and therefore make simplifications complicated.
These operators are hermitian and using this property, we let the left operator act on the left basis function and the right on the right basis function to receive the integral
\begin{equation}
{H}^\text{ZORA}_{\text{d},\text{I},\lambda\rho} 
=\int\text{d}\vec{r}\parantheses{\vec{\Pauli}\cdot\mom_\lambda}^*\omega_{\text{d},\text{I}}(\pos)\vec{\Pauli}\cdot\vec{\efield}(\pos)\parantheses{\vec{\Pauli}\cdot\mom_\rho},
\end{equation} 
where the momentum operator acting on the basis function was simplified as $\momop\varphi_\lambda=\mom_\lambda$, which is no longer an operator. 
Hence the commutators of all appearing elements in the above product are zero: 
$\left[\left(\vec{\Pauli}\cdot\mom_\rho\right),\left(\vec{\Pauli}\cdot\vec{\mathcal{E}}\right)\right]_-=\left[\omega_{\text{d},\text{I}}(\pos),\left(\vec{\Pauli}\cdot\vec{\mathcal{E}}\right)\right]_-=\left[\omega_{\text{d},\text{I}}(\pos),\left(\vec{\Pauli}\cdot\mom_\rho\right)\right]_-=0$. 
Using the Dirac relation 
\begin{equation}
\parantheses{\vec{\pauli}\cdot\vec{v}}\parantheses{\vec{\pauli}\cdot\vec{u}}
=\vec{v}\cdot\vec{u}\unity+\i\vec{\pauli}\cdot\parantheses{\vec{v}\times\vec{u}}
\end{equation}
twice (for spin-independent $\vec{u}$ and $\vec{v}$), the interaction matrix element is divided in three parts, that is
\begin{multline}
{H}^\text{ZORA}_{\text{d},\text{I},\lambda\rho}
=\int\text{d}\vec{r}\left[
\i\parantheses{\vec{p}_\lambda}^* \omega_{\text{d},\text{I}}(\pos) \cdot \left(\vec{\mathcal{E}}(\vec{r})\times\vec{p}_\rho\right)\unity
\right.\\\left.
+\left(\vec{\Pauli}\cdot\vec{p}_\lambda\right)^*\omega_{\text{d},\text{I}}(\pos)\vec{\mathcal{E}}(\vec{r})\cdot\vec{p}_\rho 
\right.\\\left.
- \vec{\Pauli}\cdot \left(\parantheses{\vec{p}_\lambda}^* \omega_{\text{d},\text{I}}(\pos)\times\left(\vec{\mathcal{E}}(\vec{r})\times\vec{p}_\rho\right)\right)
\right].
\label{eq: after2dirac}
\end{multline}
This expression can be further simplified using the Gra\ss{}mann identity $\vec{a}\times(\vec{b}\times\vec{c})=\vec{b}(\vec{a}\cdot\vec{c})-\vec{c}(\vec{a}\cdot\vec{b})$, receiving
\begin{multline}
{H}^\text{ZORA}_{\text{d},\text{I},\lambda\rho}=\hbar^2
\int\text{d}\pos\brackets{\i\nabla^*_\lambda \omega_{\text{d},\text{I}}(\pos)\cdot\parantheses{\vec{\efield}(\pos)\times\nabla_\rho}\right.\\\left.
+\parantheses{\vec{\Pauli}\cdot\nabla^*_\lambda}\omega_{\text{d},\text{I}}(\pos)\parantheses{\vec{\efield}(\pos)\cdot\nabla_\rho}\right.\\\left.
-\nabla^*_\lambda\omega_{\text{d},\text{I}}(\pos)\parantheses{\vec{\Pauli}\cdot\vec{\efield}(\pos)}\nabla_\rho\right.\\\left.
+\parantheses{\nabla^*_\lambda\cdot\vec{\efield}(\pos)}\omega_{\text{d},\text{I}}(\pos)\parantheses{\vec{\Pauli}\cdot\nabla_\rho}},
\end{multline}
where the momentum operator $\momop=-\i\hbar\nabla$ was inserted and the notation 
\begin{equation}
\nabla_\lambda= \nabla \varphi_\lambda;~~~\nabla^*_\lambda= \nabla \varphi^*_\lambda
\end{equation}
was introduced. Now we can separate terms with respect to the spatial components of the spin and write the eEDM Hamiltonian in the form
\begin{equation}
\Op{O}=\Op{O}^{(0)}+\Pauli^1\Op{O}^{(1)}+\Pauli^2\Op{O}^{(2)}+\Pauli^3\Op{O}^{(3)}.
\label{eq: generaloperator}
\end{equation}
This results in spin-free matrix-elements of the ZORA eEDM Hamiltonian
 \begin{equation}
{H}^{\text{ZORA},(0)}_{\text{d},\text{I},\lambda\rho}=\i\hbar^2\int\text{d}\vec{r}\nabla^*_\lambda \omega_{\text{d},\text{I}}(\pos)\cdot\parantheses{\vec{\efield}(\pos)\times\nabla_\rho}
\label{eq: spinindephZORAI}
\end{equation}
and matrix-elements corresponding to the three spatial directions of spin $k\not=m\not=l~\wedge~k,m,l\in\braces{1,2,3}$
\begin{multline}
{H}^{\text{ZORA},(k)}_{\text{d},\text{I},\lambda\rho}=\hbar^2\int\text{d}\vec{r}\brackets{
\partial^*_{k_\lambda}\omega_{\text{d},\text{I}}(\pos)\parantheses{\vec{\efield}(\pos)\cdot\nabla_\rho} \right.\\\left.
-\nabla_\lambda^*\omega_{\text{d},\text{I}}(\pos)\efield_k(\pos)\nabla_\rho
+\parantheses{\nabla^*_\lambda\cdot\vec{\efield}(\pos)}\omega_{\text{d},\text{I}}(\pos)\partial_{k_\rho}},
\label{eq: spindephZORAI}
\end{multline} 
where the notation 
\begin{equation}
\partial_{k_\lambda}=\partial_{k}\varphi_\lambda;~~~\partial^*_{k_\lambda}=\partial_{k}\varphi^*_\lambda
\end{equation}
was introduced. 
The internal electrical field is calculated in the approximation of Eq. \prettyref{eq: efieldapprox} and in the modified eEDM ZORA factor a model potential, introduced by van W\"{u}llen,\cite{wullen:1998} is used to alleviate the gauge dependence of ZORA. 
With that the matrix elements are implemented as
\begin{widetext}
\begin{multline}
{H}^{\text{ZORA},(0)}_{\text{d},\text{I},{\lambda\rho}}=\frac{2\i \hbar^2e d_\text{e} c^2}{4\pi \epsilon_0}\left[\sum\limits_{\alpha=1}^{N_\text{nuc}}\int\text{d}\vec{r}\cdot \frac{Z_\alpha}{\left(2m_\text{e}c^2-\tilde{V}(\vec{r})\right)^2\left((x-x_\alpha)^2+(y-y_\alpha)^2+(z-z_\alpha)^2\right)^{3/2}}
\right.\\\left.
\times\parantheses{(x-x_\alpha)\left(\partial^*_{z_\lambda}\partial_{y_\rho}-\partial^*_{y_\lambda}\partial_{z_\rho}\right)
+(y-y_\alpha)\left(\partial^*_{x_\lambda}\partial_{z_\rho}-\partial^*_{z_\lambda}\partial_{x_\rho}\right)
+(z-z_\alpha))\left(\partial^*_{y_\lambda}\partial_{x_\rho}-\partial^*_{x_\lambda}\partial_{y_\rho}\right)}\right]
\label{eq: zora1eintsf}
\end{multline}
\begin{multline}
{H}^{\text{ZORA},(k)}_{\text{d},\text{I},{\lambda\rho}}=\frac{2 \hbar^2e d_\text{e} c^2}{4\pi \epsilon_0}\left[\sum\limits_{\alpha=1}^{N_\text{nuc}}\int\text{d}\vec{r}\cdot \frac{Z_\alpha}{\left(2m_\text{e}c^2-\tilde{V}(\vec{r})\right)^2\left((x-x_\alpha)^2+(y-y_\alpha)^2+(z-z_\alpha)^2\right)^{3/2}}\right.\\\left.
\times\left((\mathrm{x}_k-\mathrm{x}_{k,\alpha})\left(\partial^*_{k_\lambda}\partial_{k_\rho}
-\partial^*_{l_\lambda}\partial_{l_\rho}-\partial^*_{m_\lambda}\partial_{m_\rho}\right)
+(\mathrm{x}_{l}-\mathrm{x}_{l,\alpha})\cdot\left(\partial^*_{l_\lambda}\partial_{k_\rho}
+\partial^*_{k_\lambda}\partial_{l_\rho}\right)
+(\mathrm{x}_{m}-\mathrm{x}_{m,\alpha})
\cdot\left(\partial^*_{m_\lambda}\partial_{k_\rho}+\partial^*_{k_\lambda}\partial_{m_\rho}\right)\right)\right]
\label{eq: zora1eintsd}
\end{multline}
\end{widetext}
These integrals can not be solved analytically and therefore numerical integration on a Becke grid is used for calculations of the eEDM matrix elements.
Finally the eEDM interaction can be calculated using the corresponding density and spin-density matrices ${\bf D}^{(\mu)}$, which are obtained from the molecular orbital coefficients $C_\lambda^{(\alpha)},C_\lambda^{(\beta)}$ of orbital $\lambda$ for up- ($\alpha$) and down-spin ($\beta$)  by the formulas
\begin{subequations}
\begin{align}
D^{(0)}_{\lambda\rho}&=\Sum{i=1}{N_\text{occ}}\brackets{\parantheses{C_{\lambda i}^{(\alpha)}}^*C_{\rho i}^{(\alpha)}+\parantheses{C_{\lambda i}^{(\beta)}}^*C_{\rho i}^{(\beta)}}\\
D^{(1)}_{\lambda\rho}&=\Sum{i=1}{N_\text{occ}}\brackets{\parantheses{C_{\lambda i}^{(\alpha)}}^*C_{\rho i}^{(\beta)}+\parantheses{C_{\lambda i}^{(\beta)}}^*C_{\rho i}^{(\alpha)}}\\
D^{(2)}_{\lambda\rho}&=-\i\Sum{i=1}{N_\text{occ}}\brackets{\parantheses{C_{\lambda i}^{(\alpha)}}^*C_{\rho i}^{(\beta)}-\parantheses{C_{\lambda i}^{(\beta)}}^*C_{\rho i}^{(\alpha)}}\\
D^{(3)}_{\lambda\rho}&=\Sum{i=1}{N_\text{occ}}\brackets{\parantheses{C_{\lambda i}^{(\alpha)}}^*C_{\rho i}^{(\alpha)}-\parantheses{C_{\lambda i}^{(\beta)}}^*C_{\rho i}^{(\beta)}},
\end{align}
\end{subequations}
where the sum runs over all $N_\text{occ}$ occupied orbitals $i$.
Then the molecular expectation value of the eEDM interaction Hamiltonian reads
\begin{multline}
{H}^{\text{ZORA}}_{\text{d},\text{I}}=\Sum{\lambda,\rho}{}\brackets{-\mathfrak{Im}\braces{D^{(0)}_{\lambda\rho}}\mathfrak{Im}\braces{{H}^{\text{ZORA},(0)}_{\text{d},\text{I},{\lambda\rho}}}\right.\\\left.
+\mathfrak{Re}\braces{D^{(1)}_{\lambda\rho}}{H}^{\text{ZORA},(1)}_{\text{d},\text{I},{\lambda\rho}}+\mathfrak{Re}\braces{D^{(2)}_{\lambda\rho}}{H}^{\text{ZORA},(2)}_{\text{d},\text{I},{\lambda\rho}}\right.\\\left.
+\mathfrak{Re}\braces{D^{(3)}_{\lambda\rho}}{H}^{\text{ZORA},(3)}_{\text{d},\text{I},{\lambda\rho}}}.
\label{eq: ZORAHd}
\end{multline}
These formulas are the working equations used to calculate the expectation value of ${H}^{\text{ZORA}}_{\text{d},\text{I}}$ in first order perturbation theory. 
In the present calculations the \mbox{(spin-)}density matrices are
obtained from a self-consistent field (SCF) calculation within the
complex generalized Hartree-Fock (cGHF) or complex generalized
Kohn-Sham (cGKS)
formalism using the ZORA Hamiltonian, but could in principle be also
obtained from a more sophisticated electronic structure treatment.\par

\subsubsection{Derivation starting from Stratagem II}
With the perturbation $\Op{H}_{\text{d},\text{II}}$ the block matrix form of the molecular Dirac equation looks like
\begin{equation}
\begin{pmatrix}
\Op{V}(\pos_i)-\epsilon&c\parantheses{\spinmompot_i+\frac{2\i d_\text{e}}{e\hbar} \momop_i^2}\\
c\parantheses{\spinmompot_i-\frac{2\i d_\text{e}}{e\hbar} \momop_i^2}&\Op{V}(\pos_i)-\epsilon-2m_\text{e}c^2\\
\end{pmatrix}
\begin{pmatrix}
\phi(\pos_i)\\
\chi(\pos_i)\\
\end{pmatrix}
=
\begin{pmatrix}
\vec{0}\\
\vec{0}\\
\end{pmatrix}.
\end{equation}
Again in the following the electronic index will be dropped. ESC yields for the small component
\begin{equation}
\chi(\pos)=\parantheses{2m_\text{e}c^2-\Op{V}(\pos)+\epsilon}^{-1}c\parantheses{\spinmompot-\frac{2\i d_\text{e}}{e\hbar} \momop^2}\phi(\pos).
\end{equation}
Now the perturbation only appears in the numerator and no special considerations respective to the denominator (in parentheses) are needed. Hence the ESC Hamiltonian can be written down as
\begin{multline}
\Op{H}^\text{ESC}_{\text{tot,II}}=\Op{V}(\pos)+c^2\parantheses{\spinmompot+\frac{2\i d_\text{e}}{e\hbar}\momop^2}\\
\times\parantheses{2m_\text{e}c^2-\Op{V}(\pos)+\epsilon}^{-1}\parantheses{\spinmompot-\frac{2\i d_\text{e}}{e\hbar}\momop^2}.
\end{multline}
Expanding the second term on the right one obtains an unperturbed ESC Hamiltonian, a perturbations linear in $d_\text{e}$ and a quadratic term in $d_\text{e}$:
\begin{widetext}
\begin{multline}
\Op{H}^\text{ESC}_{\text{tot,II}}=\Op{H}^\text{ESC}_0+\underbrace{c^2\frac{4 d_\text{e}^2}{e^2\hbar^2}\momop^2\parantheses{2m_\text{e}c^2-\Op{V}(\pos)+\epsilon}^{-1}\momop^2}_{\Op{H}^\text{ESC}_{\text{II}}(d_\text{e}^2)}\\
+\underbrace{c^2\frac{2\i d_\text{e}}{e\hbar}\momop^2\parantheses{2m_\text{e}c^2-\Op{V}(\pos)+\epsilon}^{-1}\parantheses{\spinmompot}-c^2\parantheses{\spinmompot_i}\parantheses{2m_\text{e}c^2-\Op{V}(\pos)+\epsilon}^{-1}\frac{2\i d_\text{e}}{e\hbar}\momop^2}_{\Op{H}^\text{ESC}_{\text{II}}(d_\text{e})}.
\end{multline}
\end{widetext}
When considering the terms in first order of $d_\text{e}$ only and carrying out the ZORA expansion, one obtains the resulting eEDM interaction Hamiltonian as (again with the vector potential depending terms dropped) 
\begin{equation}
\Op{H}^\text{ZORA}_{\text{d},\text{II}} =\i\momop^2\omega_{\text{d},\text{II}}(\pos)\parantheses{\spinmom}-\i\parantheses{\spinmom}\omega_{\text{d},\text{II}}(\pos)\momop^2,
\label{eq: ZORAII}
\end{equation}
where the modified ZORA factor is defined as
\begin{equation}
\omega_{\text{d},\text{II}}(\pos)=\frac{2d_\text{e}c^2}{2e\hbar m_\text{e}c^2-e\hbar\Op{V}(\vec{r})}.
\end{equation}
Its matrix elements in an arbitrary one-electron basis $\{\varphi_\lambda\}$ are
\begin{multline}
{H}^\text{ZORA}_{\text{d},\text{II},\lambda\rho} 
= \Braket{\varphi_\lambda|\Op{H}^\text{ZORA}_{\text{d},\text{II}}|\varphi_\rho}\\
=\Braket{\varphi_\lambda|\i\momop^2\omega_{\text{d},\text{II}}(\pos)\parantheses{\spinmom}-\i\parantheses{\spinmom}\omega_{\text{d},\text{II}}(\pos)\momop^2|\varphi_\rho}.
\end{multline}
Due to the simple form of the operator we can directly proceed in writing the matrix elements of $\Op{H}^\text{ZORA}_{\text{d},\text{II}} $ in terms of Eq. \prettyref{eq: generaloperator} and get in analogy to \prettyref{eq: zora1eintsd}
\begin{equation}
{H}^{\text{ZORA},(k)}_{\text{d},\text{II},\lambda\rho}=\frac{2\hbar^2d_\text{e}c^2}{e}\int\text{d}\vec{r}\frac{\partial^*_{k_\lambda}\nabla_\rho^2
+\parantheses{\nabla_{\lambda}^2}^*\partial_{k_\rho}}{2 m_\text{e}c^2-\tilde{V}(\vec{r})},
\label{eq: spindephZORAII}
\end{equation}
and there is no spin-free part in the Hamiltonian.
Again the integrals are evaluated numerically due to the appearance of the model potential $\tilde{V}(\vec{r})$ in the denominator. 
Finally, the total ZORA eEDM interaction energy derived from Stratagem II is evaluated from the (spin-)density matrices as presented in Eq. \prettyref{eq: ZORAHd} for $\Op{H}^\text{ZORA}_{\text{d},\text{I}} $.

\section{Computational Details}
For the calculation of two-component wave functions at the
GHF-/GKS-ZORA level a modified
version\cite{berger:2005a,berger:2005,isaev:2012,nahrwold:09} of the
quantum chemistry program package Turbomole\cite{ahlrichs:1989} was
used.
In order to calculate the $\mathcal{P,T}$-odd eEDM interaction, the
program was extended with the ZORA eEDM Hamiltonians implemented as
shown in Equations \prettyref{eq: zora1eintsf}, \prettyref{eq:
zora1eintsd}, \prettyref{eq: ZORAHd} and \prettyref{eq:
spindephZORAII}.\par
For density functional theory (DFT) calculations within the Kohn-Sham
framework the hybrid Becke three parameter exchange
functional and Lee, Yang and Parr correlation functional
(B3LYP)\cite{stephens:1994,vosko:1980,becke:1988,lee:1988} was employed.
For all calculations an atom centered basis set of 37~s, 34~p, 14~d and 9~f
uncontracted Gaussian functions with the exponential coefficients
$\alpha_i$ composed as an even-tempered series as $\alpha_i=a\cdot
b^{N-i};~ i=1,\dots,N$, with $b=2$ for s- and p-function and with
$b=(5/2)^{1/25}\times10^{2/5}\approx 2.6$ for d- and f-functions
was used for the heavy atom (for details see Supplementary Material).
This basis has been proven successful in calculations of nuclear-spin
dependent $\mathcal{P}$-violating interactions in heavy polar diatomic
molecules.\cite{isaev:2012,isaev:2013,isaev:2014}
The basis set centered at the fluorine atom was represented by an 
uncontracted atomic natural orbital (ANO) basis of 
triple-$\zeta$ quality\cite{roos:2004}.  
The ZORA-model potential $\tilde{V}(\pos)$ was employed with
additional damping\cite{liu:2002} as proposed by van
W\"ullen.\cite{wullen:1998}\par 
For the calculations of two-component wave functions and properties a
finite nucleus was used, described by a spherical Gaussian charge
distribution
$\rho_\alpha(\pos)=\rho_0\mathrm{e}^{-\frac{3}{2\zeta_\alpha}\pos^2}$,
where $\rho_0=eZ\parantheses{\frac{3}{2\pi\zeta_\alpha}}^{3/2}$ and
the root mean square radius $\zeta_\alpha$ of nucleus $\alpha$ was
used as suggested by Visscher and Dyall.\cite{visscher:1997}
The mass numbers $A$ were chosen as nearest integer to the standard
relative atomic mass, i.e. $^{19}$F, $^{137}$Ba, $^{173}$Yb,
$^{201}$Hg, $^{226}$Ra.\par
The nuclear distances were optimized at the levels of GHF-ZORA and
GKS-ZORA/B3LYP, respectively. For structure optimizations at
the DFT level the nucleus was approximated as a point charge. The
distances obtained are given in the results section.\par

\section{Results and Discussion}
In the following we focus on select diatomic molecules with a
$^2\Sigma_{1/2}$-ground state, which have been studied extensively in
literature\cite{kozlov:1985,dmitriev:1992,kozlov:1994,kozlov:1995,
kozlov:1997,titov:1996,mosyagin:1998,quiney:1998,
parpia:1998,kozlov:2006,nayak:2006,nayak:2006a,nayak:2007,
nayak:2009a,kudashov:2014,abe:2014,sunaga:2016,meyer:2006,
prasannaa:2015,sasmal:2016,sasmal:2016a},
namely BaF, YbF, HgF and RaF. These systems are well suited for the
validation of the here presented ZORA approach.\par
The eEDM contribution to the effective spin-rotational Hamiltonian for
diatomic molecules in a
$^2\Sigma_{1/2}$-state has the form\cite{dmitriev:1992,kozlov:1995}
\begin{equation}
H_\text{sr}=d_\text{e}W_\text{d}\Omega,
\label{eq: spinrot}
\end{equation}
where $\Omega=\vec{J_\text{e}}\cdot\vec{\lambda}$ is the projection of the total angular
momentum of the electron $\vec{J}_\text{e}$ on the molecular axis, defined by the unit
vector $\vec{\lambda}$ pointing from the heavy to the light nucleus and 
\begin{equation}
W_\text{d}=\frac{\Braket{\tilde{\psi}|\Op{H}_\text{d}|\tilde{\psi}}}{d_\text{e}\Omega},
\end{equation}
with the ZORA wave function $\tilde{\psi}$. 
In some publications, which will be referred to in the following,
instead of $W_\text{d}$ only the effective electrical field
$\efield_\text{eff}=W_\text{d}\Omega$, was reported. In the tables
below, we have converted these values for comparison then to $W_\text{d}$
using $\Omega=1/2$.\par
The GKS inter-nuclear distances are: 3.76~$a_0$ for YbF, 3.33~$a_0$ for HgH,
3.91~$a_0$ for HgF, 4.11~$a_0$ for BaF and 4.26~$a_0$ for RaF; and
$\Omega$ was in the GKS framework: 0.473 for YbF, 0.497 for HgF and
0.500 for BaF and RaF.
The GHF inter-nuclear distances are: 3.90~$a_0$ for YbF, 3.30~$a_0$ for HgH,
3.82~$a_0$ for HgF, 4.16~$a_0$ for BaF and 4.30~$a_0$ for RaF; and 
$\Omega$ was in the GHF framework: 0.498 for HgF and 0.500 for
YbF,BaF and RaF. 
\subsection{eEDM enhancement in the $^2\Sigma_{1/2}$-ground states of BaF and RaF}
We start our discussion with BaF, which was studied already in the
beginnings of the search for molecular
$\mathcal{P,T}$-violation.\cite{kozlov:1985} 
A number of calculations of $W_\text{d}$ or $W_\text{d}\Omega$ to compare
to exists for this open-shell diatomic molecule in the
literature.\par 
From \prettyref{tab: validBa} we see that the results
calculated with the two different ZORA operators are equal within the given
precision.
This suggests that differences in the transformations made to obtain 
$H_\text{d}$ are of only minor importance within ZORA.\par 
Early calculations with the generalized relativistic effective core
potential (GRECP) method without effective operator (EO) based perturbative
corrections were significantly lower in magnitude than Kozlov's
semi-empirical estimates. Changes upon inclusion of EO based corrections
imply that the inclusion of
spin-polarization is crucial for a good description of
$\mathcal{P,T}$-odd properties.\par
The most recent four-component Dirac--(Hartree)--Fock (DF) based restricted 
active space (RAS) configuration interaction (DF-RASCI) calculations are in a 
very good agreement with GRECP calculations at the highest level of theory
(RASSCF/EO, with RASSCF meaning restricted active space self-consistent field) 
and are also in good agreement with the semi-empirical
results.\cite{kozlov:1995,kozlov:1997,nayak:2006a}\par
The GHF- and GKS-ZORA calculations of this work are rather compared to
electron-correlation calculations than (paired) DF, since the complex
GHF/GKS approach already includes spin-polarization effects. 
The GHF results are in a very good agreement ($\sim 6\%$) with
previous calculations, whereas the GKS-approach underestimates
$W_\text{d}$($\sim18\%$) in magnitude. Yet, comparison of GRECP/SCF/EO 
calculations and GRECP/RASSCF/EO, which both include spin-polarization, shows
that spin-polarization effects are partially cancelled by other 
electron-correlation
effects. This trend can also be seen by comparing GKS with GHF
calculations. Thus ZORA performs amazingly well in the calculation of
a purely relativistic property such as $W_\text{d}$, although not
explicitly considering the small component.\par 
Recent publications called attention to RaF as a promising candidate
for the first measurement of $\mathcal{P}$- and $\mathcal{P,T}$-odd
effects in molecules.\cite{isaev:2013,kudashov:2014}\par
\begin{table}[h!]\centering
\begin{threeparttable}
\caption{Comparison of literature data of the $\mathcal{P,T}$-odd eEDM interaction parameter $W_\text{d}$ of the spin-rotational Hamiltonian of $^{137}$BaF calculated with different four-component methods and with a quasi-relativistic GHF/GKS-ZORA approach.}
\label{tab: validBa}
\begin{tabular}{lS[table-number-alignment=center,table-format=3.2]S[table-number-alignment=center,table-format=3.2]S[table-number-alignment=center,table-format=3.2]}
\toprule
{Method}&\multicolumn{3}{c}{$\frac{W_\text{d}~e\cdot\text{cm}}{10^{24}~\text{Hz}\cdot
h}$}\\
\midrule
Exp.+SE\tnote{1}~~ (Ref. \onlinecite{knight:1971,kozlov:1995})&&-3.5&\\
Exp.+SE\tnote{1}~~ (Ref. \onlinecite{ryzlewicz:1982,kozlov:1995})&&-4.1&\\
GRECP/SCF\tnote{2a}~~ (Ref. \onlinecite{kozlov:1997})&&-2.30\\
GRECP/SCF/EO\tnote{2b}~~ (Ref. \onlinecite{kozlov:1997})&&-3.75\\
GRECP/RASSCF\tnote{2c}~~ (Ref. \onlinecite{kozlov:1997})&&-2.24\\
GRECP/RASSCF/EO\tnote{2d}~~ (Ref. \onlinecite{kozlov:1997})&&-3.64\\
DF\tnote{3a}~~ (Ref. \onlinecite{nayak:2006a})&&-2.93\\
DF-RASCI\tnote{3b}~~ (Ref. \onlinecite{nayak:2006a})&&-3.52\\
NR-MRCI\tnote{4}~~ (Ref. \onlinecite{meyer:2006})&&-2.5\\
&$W_{\text{d},\text{I}}$&
&$W_{\text{d},\text{II}}$\\
GHF-ZORA (this work)&-3.3&
&-3.3\\
GKS-ZORA/B3LYP (this work)&-2.9&
&-2.9\\
\bottomrule
\end{tabular}
\begin{tablenotes}\footnotesize 
\item[1]  Semi-empirical estimates of $W_{\text{d}}$ calculated from experimental hyperfine coupling constants.
\item[2] Generalized relativistic effective core potential two-step approach without electron-correlation calculations (a), with effective operator technique based many-body perturbation theory of second order (b), with restricted active space SCF electron-correlation calculation (c) and with both (d).
\item[3]Dirac-Fock calculation without electron-correlation (a), with electron-correlation effects on the level of restricted active space CI (b).
\item[4] Multi-reference CI calculation within a non-relativistic framework (estimated spin-orbit energy).
\end{tablenotes}
\end{threeparttable}
\end{table}
\begin{table}[h!]\centering
\begin{threeparttable}
\caption{Comparison of literature data of the $\mathcal{P,T}$-odd eEDM interaction parameter $W_\text{d}$ of the spin-rotational Hamiltonian of RaF calculated with different four-component methods and with a quasi-relativistic GHF/GKS-ZORA approach.}
\label{tab: validRa}
\begin{tabular}{lS[table-number-alignment=center,table-format=3.2]S[table-number-alignment=center,table-format=3.2]S[table-number-alignment=center,table-format=3.2]}
\toprule
{Method}&\multicolumn{3}{c}{$\frac{W_\text{d}~e\cdot\text{cm}}{10^{24}~\text{Hz}\cdot
h}$}\\
\midrule
SODCI\tnote{1}~~ (Ref. \onlinecite{kudashov:2014})&&-24.0&\\
FS-RCCSD+$\Delta_\text{basis}$+$\Delta_\text{triples}$\tnote{2}~~(Ref. \onlinecite{kudashov:2014})&&-25.6&\\
DF-CCSD\tnote{3} ~~(Ref. \onlinecite{sasmal:2016a})&&-25.4&\\
&$W_{\text{d},\text{I}}$&
&$W_{\text{d},\text{II}}$\\
GHF-ZORA (this work)&-28.0&
&-27.3\\
GKS-ZORA/B3LYP (this work)&-25.1&
&-24.4\\
\bottomrule
\end{tabular}
\begin{tablenotes}\footnotesize 
\item[1] Spin-orbit direct configuration interaction approach
\item[2] Relativistic two-component Fock-space coupled-cluster approach with single and double amplitudes (CCSD) with basis set corrections from CCSD calculations with normal and large sized basis sets and triple-cluster corrections from CCSD calculations with and without perturbative triples.
\item[3]Dirac-Fock calculation with electron-correlation effects on the level of coupled cluster with single and double excitations (c).
\end{tablenotes}
\end{threeparttable}
\end{table}

For RaF both methods, GHF and GKS, give results that are in good agreement 
with those of three types of relativistic electron-correlation calculations 
(both below 10\% deviation, see \prettyref{tab: validRa}) with
the DFT results agreeing slightly better with the
literature values (below 5\% error). Electron correlation effects as
described on the DFT level display the same trends as observed for BaF.\par
\subsection{eEDM enhancement in the $^2\Sigma_{1/2}$-ground state of YbF}
YbF is probably the best studied
molecule with respect to molecular $\mathcal{CP}$-violation.
\begin{table}[h!]\centering
\begin{threeparttable}
\caption{Comparison of literature data of the $\mathcal{P,T}$-odd eEDM interaction parameter $W_\text{d}$ of the spin-rotational Hamiltonian of YbF calculated with different methods and with a quasi-relativistic GHF/GKS-ZORA approach.}
\label{tab: validYb}
\begin{tabular}{lS[table-number-alignment=center,table-format=3.2]S[table-number-alignment=center,table-format=3.2]S[table-number-alignment=center,table-format=3.2]}
\toprule
{Method}&\multicolumn{3}{c}{$\frac{W_\text{d}~e\cdot\text{cm}}{10^{24}~\text{Hz}\cdot h}$}\\
\midrule
Exp.+SE+corr.\tnote{1}~~(Ref. \onlinecite{vanzee:1978,kozlov:1994,kozlov:1997})&&-12.6\\
GRECP/SCF\tnote{2a}  ~~(Ref. \onlinecite{titov:1996})&&-9.1&\\
GRECP/RASSCF\tnote{2b}  ~~(Ref. \onlinecite{titov:1996})&&-9.1&\\
GRECP/RASSCF/EO\tnote{2c}  ~~(Ref. \onlinecite{mosyagin:1998})&&-12.06&\\
GRECP/RASSCF/EO+4$f$\tnote{2d} ~~ (Ref. \onlinecite{mosyagin:1998})&&-12.16&\\
RDHF+CP\tnote{3} ~~ (Ref. \onlinecite{quiney:1998})&&-6.0&\\
UDF\tnote{4} ~~ (Ref. \onlinecite{parpia:1998})&&-12.03\\
DF\tnote{5a} ~~ (Ref. \onlinecite{nayak:2006})&&-9.63\\
DF-RASCI  \tnote{5b} ~~ (Ref. \onlinecite{nayak:2006}) &&-10.88\\
DF-MBPT(2)\tnote{5c} ~~ (Ref. \onlinecite{nayak:2007}) &&-10.43\\
DF-RASCI* \tnote{5d} ~~ (Ref. \onlinecite{nayak:2009a})&&-11.64\\
DF\tnote{5a} ~~ (Ref. \onlinecite{abe:2014})&&-8.80&\\
DF-CCSD\tnote{5e} ~~ (Ref. \onlinecite{abe:2014})&&-11.17&\\
NR-MRCI\tnote{6} ~~ (Ref. \onlinecite{meyer:2006})&&-21\\
&$W_{\text{d},\text{I}}$&
&$W_{\text{d},\text{II}}$\\
GHF-ZORA (this work)&-11.6 &
&-11.4\\
GKS-ZORA/B3LYP (this work)&-10.0&
&-9.9\\
\bottomrule
\end{tabular}
\begin{tablenotes}\footnotesize
\item[1]  Semi-empirical estimates of $W_{\text{d}}$ calculated from experimental hyperfine coupling constants with correction of higher spherical waves (b).
\item[2] Generalized relativistic effective core potential two-step
approach (a), with restricted active space SCF electron-correlation
calculation without (b), with (c) effective operator technique based
many-body perturbation theory of second order and (d) additional $4f$-hole corrections.
\item[3] Restricted DHF with core-polarization corrections. Since the values are approximately by a factor of two lower, it is likely that a different definition of $W_\text{d}$ was used. 
\item[4] Unrestricted Dirac-Fock all-electron calculation.
\item[5]Dirac-Fock calculation without electron-correlation (a), with electron-correlation effects on the level of restricted active space CI (b, improved calculations: d), with electron-correlation effects on the level of second order perturbation theory (c), with electron-correlation effects on the level of coupled cluster with single and double excitations (e).
\item[6] Multi-reference CI calculation within a non-relativistic framework (estimated spin-orbit energy).
\end{tablenotes}
\end{threeparttable}
\end{table}

The values of $W_\text{d}$, calculated via Eq. \prettyref{eq: stratagemI} and
via Eq. \prettyref{eq: stratagemII}  are in an excellent agreement and
deviations are smaller than one percent (see \prettyref{tab: validYb}). This 
confirms the approximate equivalence
of the two used stratagems to calculate $W_\text{d}$ within ZORA.\par
As can be seen in \prettyref{tab: validYb} there is a large
discrepancy between the literature results with deviations of up to
30\%{}. 
Whereas calculations with Kozlov's semi-empirical model predict rather
large values for the magnitude of $W_\text{d}$ (about
$12.6\times10^{24}~\frac{\text{Hz}\cdot
h}{e\cdot\text{cm}}$),\cite{kozlov:1994,kozlov:1997} early
GRECP/RASSCF calculations and Dirac--Fock calculations without
consideration of electron-correlation yield much lower absolute
values.\cite{titov:1996,nayak:2006,abe:2014} This may result from the
neglect of spin-polarization effects, which play a major role, as has
been discussed in the previous section.\par
Yet, more recent four-component electron-correlation calculations and
GRECP calculations with perturbative effective operator corrections,
which include spin-polarization, show a better agreement and can
be taken as the most reliable of
the shown literature values.
\cite{mosyagin:1998,nayak:2007,nayak:2009,abe:2014,sunaga:2016}\par 
In comparison to DF, concerning the parameter $W_\text{d}$ the deviation of
the here presented ZORA approach is below 5\% for GHF and of the order
of 10\% for GKS. \par
Restricted DF calculations\cite{quiney:1998} deviate almost by a
factor of two from these results and therefore it can be assumed
that a different definition of the effective Hamiltonian was used.
Calculations within a non-relativistic framework, reported in Ref.
\onlinecite{meyer:2006} overestimate the magnitude of
$W_\text{d}$ by about a factor of two and are not reliable (see also 
\prettyref{tab: validHg}).\par 
Again, electron correlation corrections as estimated on the DFT level
lower the absolute value of $W_\text{d}$.

\subsection{eEDM enhancement in the $^2\Sigma_{1/2}$-ground state of HgF}
Although not studied in such detail as YbF, there is a good amount of
literature on HgF as well, partially dating back to the 1980s.\cite{kozlov:1985}
 Furthermore the calculations, which will be discussed in the
following, show that mercury compounds can provide large enhancements
of $\mathcal{P}$- and $\mathcal{CP}$-violation and therefore are very
interesting to study.\par
\begin{table}[h!]\centering
\begin{threeparttable}
\caption{Comparison of literature data of the $\mathcal{P,T}$-odd eEDM interaction parameter $W_\text{d}$ of the spin-rotational Hamiltonian of HgF calculated with different methods and with a quasi-relativistic GHF/GKS-ZORA approach.}
\label{tab: validHg}
\begin{tabular}{lS[table-number-alignment=center,table-format=3.2]S[table-number-alignment=center,table-format=3.2]S[table-number-alignment=center,table-format=3.2]}
\toprule
{Method}&\multicolumn{3}{c}{$\frac{W_\text{d}~e\cdot\text{cm}}{10^{24}~\text{Hz}\cdot h}$}\\
\midrule
Exp.+SE\tnote{1}~~ (Ref. \onlinecite{knight:1981,kozlov:1985,dmitriev:1992})
&&-47&\\
GRECP/RASSCF\tnote{2} ~~(Ref. \onlinecite{dmitriev:1992})&&-48&\\
NR-MRCI\tnote{3}~~ (Ref. \onlinecite{meyer:2006})&&-33\\
DF\tnote{4a}~~ (Ref. \onlinecite{prasannaa:2015})&&-50.42\\
DF-CCSD\tnote{4b}~~ (Ref. \onlinecite{prasannaa:2015})&&-55.81\\
&$W_{\text{d},\text{I}}$&
&$W_{\text{d},\text{II}}$\\
GHF-ZORA (this work)&-66.4&
&-65.1\\
GKS-ZORA/B3LYP (this work)&-51.1&
&-50.1\\
\bottomrule
\end{tabular}
\begin{tablenotes}\footnotesize
\item[1]Semi-empirical estimates of $W_{\text{d}}$ calculated from experimental hyperfine coupling constants.
\item[2]Generalized relativistic effective core potential two-step approach, with restricted active space SCF electron-correlation calculation without (a), with (b) effective operator technique based many-body perturbation theory of second order and (c) additional $4f$-hole corrections .
\item[3]Multi-reference CI calculation within a non-relativistic framework (estimated spin-orbit energy).
\item[4]Dirac-Fock calculation without electron-correlation (a),  with electron-correlation effects on the level of coupled cluster with single and double excitations (b).

\end{tablenotes}
\end{threeparttable}
\end{table}

As for the other compounds discussed the agreement between
$W_{\text{d},\text{I}}$ and $W_{\text{d},\text{II}}$ is excellent,
confirming the validity of the two transformations of the eEDM
Hamiltonian in the ZORA picture.\par
Whereas GRECP/RASSCF and DF results are in line with the
semi-empirical estimates by
Kozlov, although spin-polarization effects are not accounted
for \cite{dmitriev:1992,prasannaa:2015},
more recent four-component relativistic coupled cluster calculations
predict about 10\%{} larger absolute values for $W_\text{d}$.\par
As opposed to the trends observed for the molecules discussed before,
for HgF the GHF-ZORA approach
overestimates the magnitude of the results from four-component relativistic
electron-correlation calculations by more than 15\%. This appears to be 
caused by a very pronounced energetic splitting in the Kramers pair structure 
of the valence orbitals
below the singly occupied orbital, which have $\sigma$-symmetry. This
splitting is much smaller in YbF, RaF and BaF.\par
The GKS-ZORA results instead appear to be much closer to the literature data
(about 8\% deviation from DF-CCSD and even less from DF or
GRECP/RASSCF). 
Here the additional electron-correlation effects as estimated on the DFT
level, which lead to a reduction of the absolute value of $W_\text{d}$ in 
comparison to GHF, play a much more important role than for YbF within the 
GHF/GKS-ZORA approach and the difference between the GHF and GKS results is
much larger for HgF.\par
In four-component calculations the electron correlation
effects seem to be less pronounced, although accounting for spin-polarization
leads to very different results when one compares results of
GRECP/RASSCF with those of DF-CCSD.
This may be caused by partial cancellation of spin-polarization with other
correlation effects.

\section{Conclusion}
In this paper we derived a ZORA-based perturbation Hamiltonian for the
description of $\mathcal{P,T}$-odd interactions due to an electron
electric dipole moment in molecules. With calculations of promising
candidates for a search of an eEDM, we could show that a
quasi-relativistic ZORA approach is well suited for the calculation of a
purely relativistic effect, although the small component of the wave
function is not considered explicitly. The accuracy for prediction of
$W_\mathrm{d}$ is estimated to be on the order of about 20~\% for the
$^{2}\Sigma_{1/2}$-ground state molecules studied herein, if one considers
recent results obtained with electron correlation approaches as a
benchmark. This level of accuracy is presently fully sufficient for
the identification of molecular candidate systems for an eEDM search.
\par
With the quasi-relativistic approach presented in this work an
efficient calculation of the eEDM enhancement in molecules is
possible.
In future work we will study a larger number of molecules with this
approach in order to achieve a deeper understanding of the mechanisms
that lead to sizeable $\mathcal{P,T}$-odd properties in molecules.

\begin{acknowledgments}
We thank Timur Isaev for discussions. Financial support by the 
State Initiative for the Development of Scientific and
Economic Excellence (LOEWE) in the LOEWE-Focus ELCH and computer time
by the center for scientific computing (CSC) Frankfurt are gratefully
acknowledged.
\end{acknowledgments}


\begin{thebibliography}{76}%
\makeatletter
\providecommand \@ifxundefined [1]{%
 \@ifx{#1\undefined}
}%
\providecommand \@ifnum [1]{%
 \ifnum #1\expandafter \@firstoftwo
 \else \expandafter \@secondoftwo
 \fi
}%
\providecommand \@ifx [1]{%
 \ifx #1\expandafter \@firstoftwo
 \else \expandafter \@secondoftwo
 \fi
}%
\providecommand \natexlab [1]{#1}%
\providecommand \enquote  [1]{``#1''}%
\providecommand \bibnamefont  [1]{#1}%
\providecommand \bibfnamefont [1]{#1}%
\providecommand \citenamefont [1]{#1}%
\providecommand \href@noop [0]{\@secondoftwo}%
\providecommand \href [0]{\begingroup \@sanitize@url \@href}%
\providecommand \@href[1]{\@@startlink{#1}\@@href}%
\providecommand \@@href[1]{\endgroup#1\@@endlink}%
\providecommand \@sanitize@url [0]{\catcode `\\12\catcode `\$12\catcode
  `\&12\catcode `\#12\catcode `\^12\catcode `\_12\catcode `\%12\relax}%
\providecommand \@@startlink[1]{}%
\providecommand \@@endlink[0]{}%
\providecommand \url  [0]{\begingroup\@sanitize@url \@url }%
\providecommand \@url [1]{\endgroup\@href {#1}{\urlprefix }}%
\providecommand \urlprefix  [0]{URL }%
\providecommand \Eprint [0]{\href }%
\providecommand \doibase [0]{http://dx.doi.org/}%
\providecommand \selectlanguage [0]{\@gobble}%
\providecommand \bibinfo  [0]{\@secondoftwo}%
\providecommand \bibfield  [0]{\@secondoftwo}%
\providecommand \translation [1]{[#1]}%
\providecommand \BibitemOpen [0]{}%
\providecommand \bibitemStop [0]{}%
\providecommand \bibitemNoStop [0]{.\EOS\space}%
\providecommand \EOS [0]{\spacefactor3000\relax}%
\providecommand \BibitemShut  [1]{\csname bibitem#1\endcsname}%
\let\auto@bib@innerbib\@empty
\bibitem [{\citenamefont {Gross}(1996)}]{gross:1996}%
  \BibitemOpen
  \bibfield  {author} {\bibinfo {author} {\bibfnamefont {D.~J.}\ \bibnamefont
  {Gross}},\ }\href {http://www.pnas.org/content/93/25/14256.full} {\bibfield
  {journal} {\bibinfo  {journal} {Proc. Natl. Acad. Sci. USA}\ }\textbf
  {\bibinfo {volume} {93}},\ \bibinfo {pages} {14256} (\bibinfo {year}
  {1996})}\BibitemShut {NoStop}%
\bibitem [{\citenamefont {Fortson}, \citenamefont {Sandars},\ and\
  \citenamefont {Barr}(2003)}]{fortson:2003}%
  \BibitemOpen
  \bibfield  {author} {\bibinfo {author} {\bibfnamefont {N.}~\bibnamefont
  {Fortson}}, \bibinfo {author} {\bibfnamefont {P.}~\bibnamefont {Sandars}}, \
  and\ \bibinfo {author} {\bibfnamefont {S.}~\bibnamefont {Barr}},\ }\href
  {\doibase 10.1063/1.1595052} {\bibfield  {journal} {\bibinfo  {journal}
  {Phys. Today}\ }\textbf {\bibinfo {volume} {56}},\ \bibinfo {pages} {33}
  (\bibinfo {year} {2003})},\ \Eprint {http://arxiv.org/abs/arXiv:1110.0154v1}
  {arXiv:arXiv:1110.0154v1} \BibitemShut {NoStop}%
\bibitem [{\citenamefont {Schwinger}(1951)}]{schwinger:1951}%
  \BibitemOpen
  \bibfield  {author} {\bibinfo {author} {\bibfnamefont {J.}~\bibnamefont
  {Schwinger}},\ }\href {http://link.aps.org/doi/10.1103/PhysRev.82.914}
  {\bibfield  {journal} {\bibinfo  {journal} {Phys. Rev.}\ }\textbf {\bibinfo
  {volume} {82}},\ \bibinfo {pages} {914} (\bibinfo {year} {1951})}\BibitemShut
  {NoStop}%
\bibitem [{\citenamefont {Pauli}, \citenamefont {Weisskopf},\ and\
  \citenamefont {Rosenfeld}(1955)}]{pauli:1955}%
  \BibitemOpen
  \bibinfo {editor} {\bibfnamefont {W.}~\bibnamefont {Pauli}}, \bibinfo
  {editor} {\bibfnamefont {V.}~\bibnamefont {Weisskopf}}, \ and\ \bibinfo
  {editor} {\bibfnamefont {L.}~\bibnamefont {Rosenfeld}},\ eds.,\ \href
  {http://www.amazon.com/Niels-Bohr-Development-Physics-Seventieth/dp/B000KUQL76/ref=sr_1_3?ie=UTF8&#38;s=books&#38;qid=1205148170&#38;sr=1-3}
  {\emph {\bibinfo {title} {{Niels Bohr and the Development of Physics}}}}\
  (\bibinfo  {publisher} {McGraw-Hill},\ \bibinfo {year} {1955})\BibitemShut
  {NoStop}%
\bibitem [{\citenamefont {L{\"{u}}ders}(1954)}]{luders:1954}%
  \BibitemOpen
  \bibfield  {author} {\bibinfo {author} {\bibfnamefont {G.}~\bibnamefont
  {L{\"{u}}ders}},\ }\href {https://cds.cern.ch/record/1071765} {\bibfield
  {journal} {\bibinfo  {journal} {Dan. Mat. Fys. Medd.}\ }\textbf {\bibinfo
  {volume} {28}},\ \bibinfo {pages} {1} (\bibinfo {year} {1954})}\BibitemShut
  {NoStop}%
\bibitem [{\citenamefont {Khriplovich}\ and\ \citenamefont
  {Lamoreaux}(1997)}]{khriplovich:1997}%
  \BibitemOpen
  \bibfield  {author} {\bibinfo {author} {\bibfnamefont {I.~B.}\ \bibnamefont
  {Khriplovich}}\ and\ \bibinfo {author} {\bibfnamefont {S.~K.}\ \bibnamefont
  {Lamoreaux}},\ }\href@noop {} {\emph {\bibinfo {title} {\emph{CP} Violation
  without Strangeness}}}\ (\bibinfo  {publisher} {Springer},\ \bibinfo
  {address} {Berlin},\ \bibinfo {year} {1997})\BibitemShut {NoStop}%
\bibitem [{\citenamefont {Christenson}\ \emph {et~al.}(1964)\citenamefont
  {Christenson}, \citenamefont {Cronin}, \citenamefont {Fitch},\ and\
  \citenamefont {Turlay}}]{christenson:1964}%
  \BibitemOpen
  \bibfield  {author} {\bibinfo {author} {\bibfnamefont {J.~H.}\ \bibnamefont
  {Christenson}}, \bibinfo {author} {\bibfnamefont {J.~W.}\ \bibnamefont
  {Cronin}}, \bibinfo {author} {\bibfnamefont {V.~L.}\ \bibnamefont {Fitch}}, \
  and\ \bibinfo {author} {\bibfnamefont {R.}~\bibnamefont {Turlay}},\
  }\href@noop {} {\bibfield  {journal} {\bibinfo  {journal} {Phys. Rev. Lett.}\
  }\textbf {\bibinfo {volume} {13}},\ \bibinfo {pages} {138} (\bibinfo {year}
  {1964})}\BibitemShut {NoStop}%
\bibitem [{\citenamefont {Salpeter}(1958)}]{salpeter:1958}%
  \BibitemOpen
  \bibfield  {author} {\bibinfo {author} {\bibfnamefont {E.}~\bibnamefont
  {Salpeter}},\ }\href
  {http://journals.aps.org/pr/abstract/10.1103/PhysRev.112.1642} {\bibfield
  {journal} {\bibinfo  {journal} {Phys. Rev.}\ }\textbf {\bibinfo {volume}
  {112}},\ \bibinfo {pages} {1642} (\bibinfo {year} {1958})}\BibitemShut
  {NoStop}%
\bibitem [{\citenamefont {Sandars}(1965)}]{sandars:1965}%
  \BibitemOpen
  \bibfield  {author} {\bibinfo {author} {\bibfnamefont {P.}~\bibnamefont
  {Sandars}},\ }\href {\doibase 10.1016/0031-9163(65)90583-4} {\bibfield
  {journal} {\bibinfo  {journal} {Phys. Lett.}\ }\textbf {\bibinfo {volume}
  {14}},\ \bibinfo {pages} {194} (\bibinfo {year} {1965})}\BibitemShut
  {NoStop}%
\bibitem [{\citenamefont {Sandars}(1966)}]{sandars:1966}%
  \BibitemOpen
  \bibfield  {author} {\bibinfo {author} {\bibfnamefont {P.~G.~H.}\
  \bibnamefont {Sandars}},\ }\href {\doibase 10.1016/0031-9163(66)90618-4}
  {\bibfield  {journal} {\bibinfo  {journal} {Phys. Lett.}\ }\textbf {\bibinfo
  {volume} {22}},\ \bibinfo {pages} {290} (\bibinfo {year} {1966})}\BibitemShut
  {NoStop}%
\bibitem [{\citenamefont {Sandars}(1968{\natexlab{a}})}]{sandars:1968}%
  \BibitemOpen
  \bibfield  {author} {\bibinfo {author} {\bibfnamefont {P.~G.~H.}\
  \bibnamefont {Sandars}},\ }\href {\doibase 10.1088/0022-3700/1/3/326}
  {\bibfield  {journal} {\bibinfo  {journal} {J. Phys. B At. Mol. Phys.}\
  }\textbf {\bibinfo {volume} {1}},\ \bibinfo {pages} {511} (\bibinfo {year}
  {1968}{\natexlab{a}})}\BibitemShut {NoStop}%
\bibitem [{\citenamefont {Sandars}(1968{\natexlab{b}})}]{sandars:1968a}%
  \BibitemOpen
  \bibfield  {author} {\bibinfo {author} {\bibfnamefont {P.~G.~H.}\
  \bibnamefont {Sandars}},\ }\href {\doibase 10.1088/0022-3700/1/3/325}
  {\bibfield  {journal} {\bibinfo  {journal} {J. Phys. B At. Mol. Phys.}\
  }\textbf {\bibinfo {volume} {1}},\ \bibinfo {pages} {499} (\bibinfo {year}
  {1968}{\natexlab{b}})}\BibitemShut {NoStop}%
\bibitem [{\citenamefont {Sandars}(1967)}]{sandars:1967}%
  \BibitemOpen
  \bibfield  {author} {\bibinfo {author} {\bibfnamefont {P.~G.~H.}\
  \bibnamefont {Sandars}},\ }\href {\doibase 10.1103/PhysRevLett.19.1396}
  {\bibfield  {journal} {\bibinfo  {journal} {Phys. Rev. Lett.}\ }\textbf
  {\bibinfo {volume} {19}},\ \bibinfo {pages} {1396} (\bibinfo {year}
  {1967})}\BibitemShut {NoStop}%
\bibitem [{\citenamefont {Labzowsky}(1978)}]{labzowsky:1978}%
  \BibitemOpen
  \bibfield  {author} {\bibinfo {author} {\bibfnamefont {L.~N.}\ \bibnamefont
  {Labzowsky}},\ }\href@noop {} {\bibfield  {journal} {\bibinfo  {journal}
  {Sov. Phys. JETP}\ }\textbf {\bibinfo {volume} {48}},\ \bibinfo {pages} {434}
  (\bibinfo {year} {1978})}\BibitemShut {NoStop}%
\bibitem [{\citenamefont {Gorshkov}, \citenamefont {Labzowsky},\ and\
  \citenamefont {Moskalev}(1982)}]{gorshkov:1979}%
  \BibitemOpen
  \bibfield  {author} {\bibinfo {author} {\bibfnamefont {V.~G.}\ \bibnamefont
  {Gorshkov}}, \bibinfo {author} {\bibfnamefont {L.~N.}\ \bibnamefont
  {Labzowsky}}, \ and\ \bibinfo {author} {\bibfnamefont {A.~N.}\ \bibnamefont
  {Moskalev}},\ }\href@noop {} {\bibfield  {journal} {\bibinfo  {journal} {Sov.
  Phys. JETP}\ }\textbf {\bibinfo {volume} {55}},\ \bibinfo {pages} {1042}
  (\bibinfo {year} {1982})}\BibitemShut {NoStop}%
\bibitem [{\citenamefont {Sushkov}\ and\ \citenamefont
  {Flambaum}(1978)}]{sushkov:1978}%
  \BibitemOpen
  \bibfield  {author} {\bibinfo {author} {\bibfnamefont {O.~P.}\ \bibnamefont
  {Sushkov}}\ and\ \bibinfo {author} {\bibfnamefont {V.~V.}\ \bibnamefont
  {Flambaum}},\ }\href@noop {} {\bibfield  {journal} {\bibinfo  {journal} {Sov.
  Phys. JETP}\ }\textbf {\bibinfo {volume} {48}},\ \bibinfo {pages} {608}
  (\bibinfo {year} {1978})}\BibitemShut {NoStop}%
\bibitem [{\citenamefont {Sushkov}, \citenamefont {Flambaum},\ and\
  \citenamefont {Khriplovich}(1984)}]{sushkov:1984}%
  \BibitemOpen
  \bibfield  {author} {\bibinfo {author} {\bibfnamefont {O.~P.}\ \bibnamefont
  {Sushkov}}, \bibinfo {author} {\bibfnamefont {V.~V.}\ \bibnamefont
  {Flambaum}}, \ and\ \bibinfo {author} {\bibfnamefont {I.~B.}\ \bibnamefont
  {Khriplovich}},\ }\href {http://www.jetp.ac.ru/cgi-bin/dn/e_060_05_0873.pdf}
  {\bibfield  {journal} {\bibinfo  {journal} {Sov. Phys. JETP}\ }\textbf
  {\bibinfo {volume} {60}},\ \bibinfo {pages} {873} (\bibinfo {year}
  {1984})}\BibitemShut {NoStop}%
\bibitem [{\citenamefont {Flambaum}\ and\ \citenamefont
  {Khriplovich}(1985)}]{flambaum:1985}%
  \BibitemOpen
  \bibfield  {author} {\bibinfo {author} {\bibfnamefont {V.~V.}\ \bibnamefont
  {Flambaum}}\ and\ \bibinfo {author} {\bibfnamefont {I.~B.}\ \bibnamefont
  {Khriplovich}},\ }\href@noop {} {\bibfield  {journal} {\bibinfo  {journal}
  {Phys. Lett. A}\ }\textbf {\bibinfo {volume} {110}},\ \bibinfo {pages} {121}
  (\bibinfo {year} {1985})}\BibitemShut {NoStop}%
\bibitem [{\citenamefont {Kozlov}(1985)}]{kozlov:1985}%
  \BibitemOpen
  \bibfield  {author} {\bibinfo {author} {\bibfnamefont {M.~G.}\ \bibnamefont
  {Kozlov}},\ }\href@noop {} {\bibfield  {journal} {\bibinfo  {journal} {Sov.
  Phys. JETP}\ }\textbf {\bibinfo {volume} {62}},\ \bibinfo {pages} {1114}
  (\bibinfo {year} {1985})}\BibitemShut {NoStop}%
\bibitem [{\citenamefont {Baron}\ \emph {et~al.}(2014)\citenamefont {Baron},
  \citenamefont {Campbell}, \citenamefont {DeMille}, \citenamefont {Doyle},
  \citenamefont {Gabrielse}, \citenamefont {Gurevich}, \citenamefont {Hess},
  \citenamefont {Hutzler}, \citenamefont {Kirilov}, \citenamefont {Kozyryev},
  \citenamefont {O'Leary}, \citenamefont {Panda}, \citenamefont {Parsons},
  \citenamefont {Petrik}, \citenamefont {Spaun}, \citenamefont {Vutha},\ and\
  \citenamefont {West}}]{baron:2014}%
  \BibitemOpen
  \bibfield  {author} {\bibinfo {author} {\bibfnamefont {J.}~\bibnamefont
  {Baron}}, \bibinfo {author} {\bibfnamefont {W.~C.}\ \bibnamefont {Campbell}},
  \bibinfo {author} {\bibfnamefont {D.}~\bibnamefont {DeMille}}, \bibinfo
  {author} {\bibfnamefont {J.~M.}\ \bibnamefont {Doyle}}, \bibinfo {author}
  {\bibfnamefont {G.}~\bibnamefont {Gabrielse}}, \bibinfo {author}
  {\bibfnamefont {Y.~V.}\ \bibnamefont {Gurevich}}, \bibinfo {author}
  {\bibfnamefont {P.~W.}\ \bibnamefont {Hess}}, \bibinfo {author}
  {\bibfnamefont {N.~R.}\ \bibnamefont {Hutzler}}, \bibinfo {author}
  {\bibfnamefont {E.}~\bibnamefont {Kirilov}}, \bibinfo {author} {\bibfnamefont
  {I.}~\bibnamefont {Kozyryev}}, \bibinfo {author} {\bibfnamefont {B.~R.}\
  \bibnamefont {O'Leary}}, \bibinfo {author} {\bibfnamefont {C.~D.}\
  \bibnamefont {Panda}}, \bibinfo {author} {\bibfnamefont {M.~F.}\ \bibnamefont
  {Parsons}}, \bibinfo {author} {\bibfnamefont {E.~S.}\ \bibnamefont {Petrik}},
  \bibinfo {author} {\bibfnamefont {B.}~\bibnamefont {Spaun}}, \bibinfo
  {author} {\bibfnamefont {A.~C.}\ \bibnamefont {Vutha}}, \ and\ \bibinfo
  {author} {\bibfnamefont {A.~D.}\ \bibnamefont {West}},\ }\href {\doibase
  10.1126/science.1248213} {\bibfield  {journal} {\bibinfo  {journal} {Science
  (80-. ).}\ }\textbf {\bibinfo {volume} {343}},\ \bibinfo {pages} {269}
  (\bibinfo {year} {2014})},\ \Eprint {http://arxiv.org/abs/1310.7534}
  {arXiv:1310.7534} \BibitemShut {NoStop}%
\bibitem [{\citenamefont {Bernreuther}\ and\ \citenamefont
  {Suzuki}(1991)}]{bernreuther:1991}%
  \BibitemOpen
  \bibfield  {author} {\bibinfo {author} {\bibfnamefont {W.}~\bibnamefont
  {Bernreuther}}\ and\ \bibinfo {author} {\bibfnamefont {M.}~\bibnamefont
  {Suzuki}},\ }\href {\doibase 10.1103/RevModPhys.63.313} {\bibfield  {journal}
  {\bibinfo  {journal} {Rev. Mod. Phys.}\ }\textbf {\bibinfo {volume} {63}},\
  \bibinfo {pages} {313} (\bibinfo {year} {1991})}\BibitemShut {NoStop}%
\bibitem [{\citenamefont {Nir}(2000)}]{nir:2000}%
  \BibitemOpen
  \bibfield  {author} {\bibinfo {author} {\bibfnamefont {Y.}~\bibnamefont
  {Nir}},\ }in\ \href {http://adsabs.harvard.edu/abs/2000paph.conf..165N}
  {\emph {\bibinfo {booktitle} {Part. Phys.}}},\ \bibinfo {editor} {edited by\
  \bibinfo {editor} {\bibfnamefont {G.}~\bibnamefont {Senjanovi{\'{c}}}}\ and\
  \bibinfo {editor} {\bibfnamefont {A.~Y.}\ \bibnamefont {Smirnov}}}\ (\bibinfo
  {year} {2000})\ p.\ \bibinfo {pages} {165}\BibitemShut {NoStop}%
\bibitem [{\citenamefont {Kudashov}\ \emph {et~al.}(2014)\citenamefont
  {Kudashov}, \citenamefont {Petrov}, \citenamefont {Skripnikov}, \citenamefont
  {Mosyagin}, \citenamefont {Isaev}, \citenamefont {Berger},\ and\
  \citenamefont {Titov}}]{kudashov:2014}%
  \BibitemOpen
  \bibfield  {author} {\bibinfo {author} {\bibfnamefont {A.~D.}\ \bibnamefont
  {Kudashov}}, \bibinfo {author} {\bibfnamefont {A.~N.}\ \bibnamefont
  {Petrov}}, \bibinfo {author} {\bibfnamefont {L.~V.}\ \bibnamefont
  {Skripnikov}}, \bibinfo {author} {\bibfnamefont {N.~S.}\ \bibnamefont
  {Mosyagin}}, \bibinfo {author} {\bibfnamefont {T.~A.}\ \bibnamefont {Isaev}},
  \bibinfo {author} {\bibfnamefont {R.}~\bibnamefont {Berger}}, \ and\ \bibinfo
  {author} {\bibfnamefont {A.~V.}\ \bibnamefont {Titov}},\ }\href {\doibase
  10.1103/PhysRevA.90.052513} {\bibfield  {journal} {\bibinfo  {journal} {Phys.
  Rev. A}\ }\textbf {\bibinfo {volume} {90}},\ \bibinfo {pages} {052513}
  (\bibinfo {year} {2014})}\BibitemShut {NoStop}%
\bibitem [{\citenamefont {Skripnikov}\ and\ \citenamefont
  {Titov}(2015)}]{skripnikov:2015}%
  \BibitemOpen
  \bibfield  {author} {\bibinfo {author} {\bibfnamefont {L.~V.}\ \bibnamefont
  {Skripnikov}}\ and\ \bibinfo {author} {\bibfnamefont {A.~V.}\ \bibnamefont
  {Titov}},\ }\href {\doibase 10.1103/PhysRevA.91.042504} {\bibfield  {journal}
  {\bibinfo  {journal} {Phys. Rev. A}\ }\textbf {\bibinfo {volume} {91}},\
  \bibinfo {pages} {042504} (\bibinfo {year} {2015})},\ \Eprint
  {http://arxiv.org/abs/arXiv:1503.01001v2} {arXiv:arXiv:1503.01001v2}
  \BibitemShut {NoStop}%
\bibitem [{\citenamefont {Prasannaa}\ \emph {et~al.}(2015)\citenamefont
  {Prasannaa}, \citenamefont {Vutha}, \citenamefont {Abe},\ and\ \citenamefont
  {Das}}]{prasannaa:2015}%
  \BibitemOpen
  \bibfield  {author} {\bibinfo {author} {\bibfnamefont {V.~S.}\ \bibnamefont
  {Prasannaa}}, \bibinfo {author} {\bibfnamefont {A.~C.}\ \bibnamefont
  {Vutha}}, \bibinfo {author} {\bibfnamefont {M.}~\bibnamefont {Abe}}, \ and\
  \bibinfo {author} {\bibfnamefont {B.~P.}\ \bibnamefont {Das}},\ }\href
  {\doibase 10.1103/PhysRevLett.114.183001} {\bibfield  {journal} {\bibinfo
  {journal} {Phys. Rev. Lett.}\ }\textbf {\bibinfo {volume} {114}},\ \bibinfo
  {pages} {183001} (\bibinfo {year} {2015})},\ \Eprint
  {http://arxiv.org/abs/1410.5138} {arXiv:1410.5138} \BibitemShut {NoStop}%
\bibitem [{\citenamefont {Denis}\ \emph {et~al.}(2015)\citenamefont {Denis},
  \citenamefont {Norby}, \citenamefont {Jensen}, \citenamefont {Gomes},
  \citenamefont {Nayak}, \citenamefont {Knecht},\ and\ \citenamefont
  {Fleig}}]{denis:2015}%
  \BibitemOpen
  \bibfield  {author} {\bibinfo {author} {\bibfnamefont {M.}~\bibnamefont
  {Denis}}, \bibinfo {author} {\bibfnamefont {M.~S.}\ \bibnamefont {Norby}},
  \bibinfo {author} {\bibfnamefont {H.~J.~A.}\ \bibnamefont {Jensen}}, \bibinfo
  {author} {\bibfnamefont {A.~S.~P.}\ \bibnamefont {Gomes}}, \bibinfo {author}
  {\bibfnamefont {M.~K.}\ \bibnamefont {Nayak}}, \bibinfo {author}
  {\bibfnamefont {S.}~\bibnamefont {Knecht}}, \ and\ \bibinfo {author}
  {\bibfnamefont {T.}~\bibnamefont {Fleig}},\ }\href {\doibase
  10.1088/1367-2630/17/4/043005} {\bibfield  {journal} {\bibinfo  {journal}
  {New J. Phys.}\ }\textbf {\bibinfo {volume} {17}},\ \bibinfo {pages} {43005}
  (\bibinfo {year} {2015})}\BibitemShut {NoStop}%
\bibitem [{\citenamefont {Fleig}, \citenamefont {Nayak},\ and\ \citenamefont
  {Kozlov}(2016)}]{fleig:2016}%
  \BibitemOpen
  \bibfield  {author} {\bibinfo {author} {\bibfnamefont {T.}~\bibnamefont
  {Fleig}}, \bibinfo {author} {\bibfnamefont {M.~K.}\ \bibnamefont {Nayak}}, \
  and\ \bibinfo {author} {\bibfnamefont {M.~G.}\ \bibnamefont {Kozlov}},\
  }\href {\doibase 10.1103/PhysRevA.93.012505} {\bibfield  {journal} {\bibinfo
  {journal} {Phys. Rev. A}\ }\textbf {\bibinfo {volume} {93}},\ \bibinfo
  {pages} {012505} (\bibinfo {year} {2016})},\ \Eprint
  {http://arxiv.org/abs/arXiv:1512.08729v1} {arXiv:arXiv:1512.08729v1}
  \BibitemShut {NoStop}%
\bibitem [{\citenamefont {Sasmal}\ \emph
  {et~al.}(2016{\natexlab{a}})\citenamefont {Sasmal}, \citenamefont {Pathak},
  \citenamefont {Nayak}, \citenamefont {Vaval},\ and\ \citenamefont
  {Pal}}]{sasmal:2016}%
  \BibitemOpen
  \bibfield  {author} {\bibinfo {author} {\bibfnamefont {S.}~\bibnamefont
  {Sasmal}}, \bibinfo {author} {\bibfnamefont {H.}~\bibnamefont {Pathak}},
  \bibinfo {author} {\bibfnamefont {M.~K.}\ \bibnamefont {Nayak}}, \bibinfo
  {author} {\bibfnamefont {N.}~\bibnamefont {Vaval}}, \ and\ \bibinfo {author}
  {\bibfnamefont {S.}~\bibnamefont {Pal}},\ }\href {\doibase 10.1063/1.4944673}
  {\bibfield  {journal} {\bibinfo  {journal} {J. Chem. Phys.}\ }\textbf
  {\bibinfo {volume} {144}},\ \bibinfo {pages} {124307} (\bibinfo {year}
  {2016}{\natexlab{a}})},\ \Eprint {http://arxiv.org/abs/arXiv:1602.08200v1}
  {arXiv:arXiv:1602.08200v1} \BibitemShut {NoStop}%
\bibitem [{\citenamefont {Sasmal}\ \emph
  {et~al.}(2016{\natexlab{b}})\citenamefont {Sasmal}, \citenamefont {Pathak},
  \citenamefont {Nayak}, \citenamefont {Vaval},\ and\ \citenamefont
  {Pal}}]{sasmal:2016a}%
  \BibitemOpen
  \bibfield  {author} {\bibinfo {author} {\bibfnamefont {S.}~\bibnamefont
  {Sasmal}}, \bibinfo {author} {\bibfnamefont {H.}~\bibnamefont {Pathak}},
  \bibinfo {author} {\bibfnamefont {M.~K.}\ \bibnamefont {Nayak}}, \bibinfo
  {author} {\bibfnamefont {N.}~\bibnamefont {Vaval}}, \ and\ \bibinfo {author}
  {\bibfnamefont {S.}~\bibnamefont {Pal}},\ }\href {\doibase
  10.1103/PhysRevA.93.062506} {\bibfield  {journal} {\bibinfo  {journal} {Phys.
  Rev. A}\ }\textbf {\bibinfo {volume} {93}},\ \bibinfo {pages} {062506}
  (\bibinfo {year} {2016}{\natexlab{b}})}\BibitemShut {NoStop}%
\bibitem [{\citenamefont {Sunaga}\ \emph {et~al.}(2016)\citenamefont {Sunaga},
  \citenamefont {Abe}, \citenamefont {Hada},\ and\ \citenamefont
  {Das}}]{sunaga:2016}%
  \BibitemOpen
  \bibfield  {author} {\bibinfo {author} {\bibfnamefont {A.}~\bibnamefont
  {Sunaga}}, \bibinfo {author} {\bibfnamefont {M.}~\bibnamefont {Abe}},
  \bibinfo {author} {\bibfnamefont {M.}~\bibnamefont {Hada}}, \ and\ \bibinfo
  {author} {\bibfnamefont {B.~P.}\ \bibnamefont {Das}},\ }\href {\doibase
  10.1103/PhysRevA.93.042507} {\bibfield  {journal} {\bibinfo  {journal} {Phys.
  Rev. A}\ }\textbf {\bibinfo {volume} {93}},\ \bibinfo {pages} {042507}
  (\bibinfo {year} {2016})}\BibitemShut {NoStop}%
\bibitem [{\citenamefont {Sunaga}\ \emph {et~al.}(2017)\citenamefont {Sunaga},
  \citenamefont {Abe}, \citenamefont {Hada},\ and\ \citenamefont
  {Das}}]{sunaga:2017}%
  \BibitemOpen
  \bibfield  {author} {\bibinfo {author} {\bibfnamefont {A.}~\bibnamefont
  {Sunaga}}, \bibinfo {author} {\bibfnamefont {M.}~\bibnamefont {Abe}},
  \bibinfo {author} {\bibfnamefont {M.}~\bibnamefont {Hada}}, \ and\ \bibinfo
  {author} {\bibfnamefont {B.~P.}\ \bibnamefont {Das}},\ }\href {\doibase
  10.1103/PhysRevA.95.012502} {\bibfield  {journal} {\bibinfo  {journal} {Phys.
  Rev. A}\ }\textbf {\bibinfo {volume} {95}},\ \bibinfo {pages} {012502}
  (\bibinfo {year} {2017})}\BibitemShut {NoStop}%
\bibitem [{\citenamefont {Dyall}\ and\ \citenamefont {{F\ae{}gri,
  Jr.}}(2007)}]{dyall:2007}%
  \BibitemOpen
  \bibfield  {author} {\bibinfo {author} {\bibfnamefont {K.~G.}\ \bibnamefont
  {Dyall}}\ and\ \bibinfo {author} {\bibfnamefont {K.}~\bibnamefont
  {{F\ae{}gri, Jr.}}},\ }\href@noop {} {\emph {\bibinfo {title} {{Introduction
  to Relativistic Quantum Chemistry}}}}\ (\bibinfo  {publisher} {Oxford
  University Press},\ \bibinfo {year} {2007})\BibitemShut {NoStop}%
\bibitem [{\citenamefont {van Leeuwen}\ \emph {et~al.}(1994)\citenamefont {van
  Leeuwen}, \citenamefont {van Lenthe}, \citenamefont {Baerends},\ and\
  \citenamefont {Snijders}}]{leeuwen:1994}%
  \BibitemOpen
  \bibfield  {author} {\bibinfo {author} {\bibfnamefont {R.}~\bibnamefont {van
  Leeuwen}}, \bibinfo {author} {\bibfnamefont {E.}~\bibnamefont {van Lenthe}},
  \bibinfo {author} {\bibfnamefont {E.-J.}\ \bibnamefont {Baerends}}, \ and\
  \bibinfo {author} {\bibfnamefont {J.~G.}\ \bibnamefont {Snijders}},\
  }\href@noop {} {\bibfield  {journal} {\bibinfo  {journal} {J. Chem. Phys.}\
  }\textbf {\bibinfo {volume} {101}},\ \bibinfo {pages} {1272} (\bibinfo {year}
  {1994})}\BibitemShut {NoStop}%
\bibitem [{\citenamefont {Berger}, \citenamefont {Langermann},\ and\
  \citenamefont {van W{\"u}llen}(2005)}]{berger:2005}%
  \BibitemOpen
  \bibfield  {author} {\bibinfo {author} {\bibfnamefont {R.}~\bibnamefont
  {Berger}}, \bibinfo {author} {\bibfnamefont {N.}~\bibnamefont {Langermann}},
  \ and\ \bibinfo {author} {\bibfnamefont {C.}~\bibnamefont {van W{\"u}llen}},\
  }\href {\doibase 10.1103/PhysRevA.71.042105} {\bibfield  {journal} {\bibinfo
  {journal} {Phys. Rev. A}\ }\textbf {\bibinfo {volume} {71}},\ \bibinfo
  {pages} {042105} (\bibinfo {year} {2005})}\BibitemShut {NoStop}%
\bibitem [{\citenamefont {Berger}\ and\ \citenamefont {van
  W{\"u}llen}(2005)}]{berger:2005a}%
  \BibitemOpen
  \bibfield  {author} {\bibinfo {author} {\bibfnamefont {R.}~\bibnamefont
  {Berger}}\ and\ \bibinfo {author} {\bibfnamefont {C.}~\bibnamefont {van
  W{\"u}llen}},\ }\href {\doibase 10.1063/1.1869467} {\bibfield  {journal}
  {\bibinfo  {journal} {J. Chem. Phys.}\ }\textbf {\bibinfo {volume} {122}},\
  \bibinfo {pages} {134316} (\bibinfo {year} {2005})}\BibitemShut {NoStop}%
\bibitem [{\citenamefont {Berger}\ and\ \citenamefont
  {Stuber}(2007)}]{berger:2007}%
  \BibitemOpen
  \bibfield  {author} {\bibinfo {author} {\bibfnamefont {R.}~\bibnamefont
  {Berger}}\ and\ \bibinfo {author} {\bibfnamefont {J.~L.}\ \bibnamefont
  {Stuber}},\ }\href {\doibase 10.1080/00268970601126759} {\bibfield  {journal}
  {\bibinfo  {journal} {Mol. Phys.}\ }\textbf {\bibinfo {volume} {105}},\
  \bibinfo {pages} {41} (\bibinfo {year} {2007})}\BibitemShut {NoStop}%
\bibitem [{\citenamefont {Berger}(2008)}]{berger:2008}%
  \BibitemOpen
  \bibfield  {author} {\bibinfo {author} {\bibfnamefont {R.}~\bibnamefont
  {Berger}},\ }\href@noop {} {\bibfield  {journal} {\bibinfo  {journal} {J.
  Chem. Phys.}\ }\textbf {\bibinfo {volume} {129}},\ \bibinfo {pages} {154105}
  (\bibinfo {year} {2008})}\BibitemShut {NoStop}%
\bibitem [{\citenamefont {Nahrwold}\ and\ \citenamefont
  {Berger}(2009)}]{nahrwold:09}%
  \BibitemOpen
  \bibfield  {author} {\bibinfo {author} {\bibfnamefont {S.}~\bibnamefont
  {Nahrwold}}\ and\ \bibinfo {author} {\bibfnamefont {R.}~\bibnamefont
  {Berger}},\ }\href {\doibase 10.1063/1.3103643} {\bibfield  {journal}
  {\bibinfo  {journal} {J. Chem. Phys.}\ }\textbf {\bibinfo {volume} {130}},\
  \bibinfo {pages} {214101} (\bibinfo {year} {2009})}\BibitemShut {NoStop}%
\bibitem [{\citenamefont {Isaev}\ and\ \citenamefont
  {Berger}(2012)}]{isaev:2012}%
  \BibitemOpen
  \bibfield  {author} {\bibinfo {author} {\bibfnamefont {T.~A.}\ \bibnamefont
  {Isaev}}\ and\ \bibinfo {author} {\bibfnamefont {R.}~\bibnamefont {Berger}},\
  }\href {\doibase 10.1103/PhysRevA.86.062515} {\bibfield  {journal} {\bibinfo
  {journal} {Phys. Rev. A}\ }\textbf {\bibinfo {volume} {86}},\ \bibinfo
  {pages} {062515} (\bibinfo {year} {2012})}\BibitemShut {NoStop}%
\bibitem [{\citenamefont {{Isaev}}\ and\ \citenamefont
  {{Berger}}(2013)}]{isaev:2013}%
  \BibitemOpen
  \bibfield  {author} {\bibinfo {author} {\bibfnamefont {T.~A.}\ \bibnamefont
  {{Isaev}}}\ and\ \bibinfo {author} {\bibfnamefont {R.}~\bibnamefont
  {{Berger}}},\ }\href@noop {} {\bibfield  {journal} {\bibinfo  {journal}
  {ArXiv e-prints}\ } (\bibinfo {year} {2013})},\ \Eprint
  {http://arxiv.org/abs/1302.5682} {arXiv:1302.5682 [physics.chem-ph]}
  \BibitemShut {NoStop}%
\bibitem [{\citenamefont {Isaev}\ and\ \citenamefont
  {Berger}(2014)}]{isaev:2014}%
  \BibitemOpen
  \bibfield  {author} {\bibinfo {author} {\bibfnamefont {T.~A.}\ \bibnamefont
  {Isaev}}\ and\ \bibinfo {author} {\bibfnamefont {R.}~\bibnamefont {Berger}},\
  }\href {\doibase 10.1016/j.jms.2014.01.014} {\bibfield  {journal} {\bibinfo
  {journal} {J. Mol. Spectrosc.}\ }\textbf {\bibinfo {volume} {300}},\ \bibinfo
  {pages} {26} (\bibinfo {year} {2014})}\BibitemShut {NoStop}%
\bibitem [{\citenamefont {Lindroth}, \citenamefont {Lynn},\ and\ \citenamefont
  {Sandars}(1989)}]{lindroth:1989}%
  \BibitemOpen
  \bibfield  {author} {\bibinfo {author} {\bibfnamefont {E.}~\bibnamefont
  {Lindroth}}, \bibinfo {author} {\bibfnamefont {B.~W.}\ \bibnamefont {Lynn}},
  \ and\ \bibinfo {author} {\bibfnamefont {P.~G.~H.}\ \bibnamefont {Sandars}},\
  }\href@noop {} {\bibfield  {journal} {\bibinfo  {journal} {J. Phys. B}\
  }\textbf {\bibinfo {volume} {22}},\ \bibinfo {pages} {559} (\bibinfo {year}
  {1989})}\BibitemShut {NoStop}%
\bibitem [{\citenamefont {Schiff}(1963)}]{schiff:1963}%
  \BibitemOpen
  \bibfield  {author} {\bibinfo {author} {\bibfnamefont {L.~I.}\ \bibnamefont
  {Schiff}},\ }\href {\doibase 10.1103/PhysRev.132.2194} {\bibfield  {journal}
  {\bibinfo  {journal} {Phys. Rev.}\ }\textbf {\bibinfo {volume} {132}},\
  \bibinfo {pages} {2194} (\bibinfo {year} {1963})}\BibitemShut {NoStop}%
\bibitem [{\citenamefont {M{\aa}rtensson-Pendrill}\ and\ \citenamefont
  {{\"{O}}ster}(1987)}]{martensson-pendrill:1987}%
  \BibitemOpen
  \bibfield  {author} {\bibinfo {author} {\bibfnamefont {A.}~\bibnamefont
  {M{\aa}rtensson-Pendrill}}\ and\ \bibinfo {author} {\bibfnamefont
  {P.}~\bibnamefont {{\"{O}}ster}},\ }\href {\doibase
  10.1088/0031-8949/36/3/011} {\bibfield  {journal} {\bibinfo  {journal} {Phys.
  Scr.}\ }\textbf {\bibinfo {volume} {444}},\ \bibinfo {pages} {444} (\bibinfo
  {year} {1987})}\BibitemShut {NoStop}%
\bibitem [{\citenamefont {Commins}\ and\ \citenamefont
  {DeMille}(2010)}]{commins:2010}%
  \BibitemOpen
  \bibfield  {author} {\bibinfo {author} {\bibfnamefont {E.~D.}\ \bibnamefont
  {Commins}}\ and\ \bibinfo {author} {\bibfnamefont {D.~P.}\ \bibnamefont
  {DeMille}},\ }in\ \href@noop {} {\emph {\bibinfo {booktitle} {Lept. Dipole
  Moments}}},\ \bibinfo {editor} {edited by\ \bibinfo {editor} {\bibfnamefont
  {B.~L.}\ \bibnamefont {Roberts}}\ and\ \bibinfo {editor} {\bibfnamefont
  {W.~J.}\ \bibnamefont {Marciano}}}\ (\bibinfo  {publisher} {World Scientific
  Publishing Company Pte Limited},\ \bibinfo {year} {2010})\ p.\ \bibinfo
  {pages} {519}\BibitemShut {NoStop}%
\bibitem [{\citenamefont {Chang}, \citenamefont {Pelissier},\ and\
  \citenamefont {Durand}(1986)}]{chang:1986}%
  \BibitemOpen
  \bibfield  {author} {\bibinfo {author} {\bibfnamefont {C.}~\bibnamefont
  {Chang}}, \bibinfo {author} {\bibfnamefont {M.}~\bibnamefont {Pelissier}}, \
  and\ \bibinfo {author} {\bibfnamefont {P.}~\bibnamefont {Durand}},\
  }\href@noop {} {\bibfield  {journal} {\bibinfo  {journal} {Phys. Scr.}\
  }\textbf {\bibinfo {volume} {34}},\ \bibinfo {pages} {394} (\bibinfo {year}
  {1986})}\BibitemShut {NoStop}%
\bibitem [{\citenamefont {van Lenthe}, \citenamefont {Baerends},\ and\
  \citenamefont {Snijders}(1993)}]{lenthe:1993}%
  \BibitemOpen
  \bibfield  {author} {\bibinfo {author} {\bibfnamefont {E.}~\bibnamefont {van
  Lenthe}}, \bibinfo {author} {\bibfnamefont {E.-J.}\ \bibnamefont {Baerends}},
  \ and\ \bibinfo {author} {\bibfnamefont {J.~G.}\ \bibnamefont {Snijders}},\
  }\href {\doibase 10.1063/1.466059} {\bibfield  {journal} {\bibinfo  {journal}
  {J. Chem. Phys.}\ }\textbf {\bibinfo {volume} {99}},\ \bibinfo {pages} {4597}
  (\bibinfo {year} {1993})}\BibitemShut {NoStop}%
\bibitem [{\citenamefont {van W{\"u}llen}(1998)}]{wullen:1998}%
  \BibitemOpen
  \bibfield  {author} {\bibinfo {author} {\bibfnamefont {C.}~\bibnamefont {van
  W{\"u}llen}},\ }\href {\doibase 10.1063/1.476576} {\bibfield  {journal}
  {\bibinfo  {journal} {J. Chem. Phys.}\ }\textbf {\bibinfo {volume} {109}},\
  \bibinfo {pages} {392} (\bibinfo {year} {1998})}\BibitemShut {NoStop}%
\bibitem [{\citenamefont {Ahlrichs}\ \emph {et~al.}(1989)\citenamefont
  {Ahlrichs}, \citenamefont {B{\"a}r}, \citenamefont {H{\"a}ser}, \citenamefont
  {Horn},\ and\ \citenamefont {K{\"o}lmel}}]{ahlrichs:1989}%
  \BibitemOpen
  \bibfield  {author} {\bibinfo {author} {\bibfnamefont {R.}~\bibnamefont
  {Ahlrichs}}, \bibinfo {author} {\bibfnamefont {M.}~\bibnamefont {B{\"a}r}},
  \bibinfo {author} {\bibfnamefont {M.}~\bibnamefont {H{\"a}ser}}, \bibinfo
  {author} {\bibfnamefont {H.}~\bibnamefont {Horn}}, \ and\ \bibinfo {author}
  {\bibfnamefont {C.}~\bibnamefont {K{\"o}lmel}},\ }\href {\doibase
  10.1016/0009-2614(89)85118-8} {\bibfield  {journal} {\bibinfo  {journal}
  {Chem. Phys. Lett.}\ }\textbf {\bibinfo {volume} {162}},\ \bibinfo {pages}
  {165} (\bibinfo {year} {1989})}\BibitemShut {NoStop}%
\bibitem [{\citenamefont {Stephens}\ \emph {et~al.}(1994)\citenamefont
  {Stephens}, \citenamefont {Devlin}, \citenamefont {Chabalowski},\ and\
  \citenamefont {Frisch}}]{stephens:1994}%
  \BibitemOpen
  \bibfield  {author} {\bibinfo {author} {\bibfnamefont {P.~J.}\ \bibnamefont
  {Stephens}}, \bibinfo {author} {\bibfnamefont {F.~J.}\ \bibnamefont
  {Devlin}}, \bibinfo {author} {\bibfnamefont {C.~F.}\ \bibnamefont
  {Chabalowski}}, \ and\ \bibinfo {author} {\bibfnamefont {M.~J.}\ \bibnamefont
  {Frisch}},\ }\href@noop {} {\bibfield  {journal} {\bibinfo  {journal} {J.
  Phys. Chem.}\ }\textbf {\bibinfo {volume} {98}},\ \bibinfo {pages} {11623}
  (\bibinfo {year} {1994})}\BibitemShut {NoStop}%
\bibitem [{\citenamefont {Vosko}, \citenamefont {Wilk},\ and\ \citenamefont
  {Nuisar}(1980)}]{vosko:1980}%
  \BibitemOpen
  \bibfield  {author} {\bibinfo {author} {\bibfnamefont {S.~H.}\ \bibnamefont
  {Vosko}}, \bibinfo {author} {\bibfnamefont {L.}~\bibnamefont {Wilk}}, \ and\
  \bibinfo {author} {\bibfnamefont {M.}~\bibnamefont {Nuisar}},\ }\href
  {\doibase 10.1139/p80-159} {\bibfield  {journal} {\bibinfo  {journal} {Can.
  J. Phys.}\ }\textbf {\bibinfo {volume} {58}},\ \bibinfo {pages} {1200}
  (\bibinfo {year} {1980})}\BibitemShut {NoStop}%
\bibitem [{\citenamefont {Becke}(1988)}]{becke:1988}%
  \BibitemOpen
  \bibfield  {author} {\bibinfo {author} {\bibfnamefont {A.~D.}\ \bibnamefont
  {Becke}},\ }\href@noop {} {\bibfield  {journal} {\bibinfo  {journal} {Phys.
  Rev. A}\ }\textbf {\bibinfo {volume} {38}},\ \bibinfo {pages} {3098}
  (\bibinfo {year} {1988})}\BibitemShut {NoStop}%
\bibitem [{\citenamefont {Lee}, \citenamefont {Yang},\ and\ \citenamefont
  {Parr}(1988)}]{lee:1988}%
  \BibitemOpen
  \bibfield  {author} {\bibinfo {author} {\bibfnamefont {C.}~\bibnamefont
  {Lee}}, \bibinfo {author} {\bibfnamefont {W.}~\bibnamefont {Yang}}, \ and\
  \bibinfo {author} {\bibfnamefont {R.~G.}\ \bibnamefont {Parr}},\ }\href
  {\doibase 10.1103/PhysRevB.37.785} {\bibfield  {journal} {\bibinfo  {journal}
  {Phys. Rev. B}\ }\textbf {\bibinfo {volume} {37}},\ \bibinfo {pages} {785}
  (\bibinfo {year} {1988})}\BibitemShut {NoStop}%
\bibitem [{\citenamefont {Roos}\ \emph {et~al.}(2004)\citenamefont {Roos},
  \citenamefont {Lindh}, \citenamefont {Malmqvist}, \citenamefont {Veryazov},\
  and\ \citenamefont {Widmark}}]{roos:2004}%
  \BibitemOpen
  \bibfield  {author} {\bibinfo {author} {\bibfnamefont {B.~O.}\ \bibnamefont
  {Roos}}, \bibinfo {author} {\bibfnamefont {R.}~\bibnamefont {Lindh}},
  \bibinfo {author} {\bibfnamefont {P.~k.}\ \bibnamefont {Malmqvist}}, \bibinfo
  {author} {\bibfnamefont {V.}~\bibnamefont {Veryazov}}, \ and\ \bibinfo
  {author} {\bibfnamefont {P.~O.}\ \bibnamefont {Widmark}},\ }\href {\doibase
  10.1021/jp031064+} {\bibfield  {journal} {\bibinfo  {journal} {J. Phys. Chem.
  A}\ }\textbf {\bibinfo {volume} {108}},\ \bibinfo {pages} {2851} (\bibinfo
  {year} {2004})}\BibitemShut {NoStop}%
\bibitem [{\citenamefont {Liu}\ \emph {et~al.}(2002)\citenamefont {Liu},
  \citenamefont {van W{\"u}llen}, \citenamefont {Wang},\ and\ \citenamefont
  {Li}}]{liu:2002}%
  \BibitemOpen
  \bibfield  {author} {\bibinfo {author} {\bibfnamefont {W.}~\bibnamefont
  {Liu}}, \bibinfo {author} {\bibfnamefont {C.}~\bibnamefont {van W{\"u}llen}},
  \bibinfo {author} {\bibfnamefont {F.}~\bibnamefont {Wang}}, \ and\ \bibinfo
  {author} {\bibfnamefont {L.}~\bibnamefont {Li}},\ }\href@noop {} {\bibfield
  {journal} {\bibinfo  {journal} {J. Chem. Phys.}\ }\textbf {\bibinfo {volume}
  {116}},\ \bibinfo {pages} {3626} (\bibinfo {year} {2002})}\BibitemShut
  {NoStop}%
\bibitem [{\citenamefont {Visscher}\ and\ \citenamefont
  {Dyall}(1997)}]{visscher:1997}%
  \BibitemOpen
  \bibfield  {author} {\bibinfo {author} {\bibfnamefont {L.}~\bibnamefont
  {Visscher}}\ and\ \bibinfo {author} {\bibfnamefont {K.~G.}\ \bibnamefont
  {Dyall}},\ }\href@noop {} {\bibfield  {journal} {\bibinfo  {journal} {At.
  Data Nucl. Data Tables}\ }\textbf {\bibinfo {volume} {67}},\ \bibinfo {pages}
  {207} (\bibinfo {year} {1997})}\BibitemShut {NoStop}%
\bibitem [{\citenamefont {Dmitriev}\ \emph {et~al.}(1992)\citenamefont
  {Dmitriev}, \citenamefont {Khait}, \citenamefont {Kozlov}, \citenamefont
  {Labzowsky}, \citenamefont {Mitrushenkov}, \citenamefont {Shtoff},\ and\
  \citenamefont {Titov}}]{dmitriev:1992}%
  \BibitemOpen
  \bibfield  {author} {\bibinfo {author} {\bibfnamefont {Y.~Y.}\ \bibnamefont
  {Dmitriev}}, \bibinfo {author} {\bibfnamefont {Y.~G.}\ \bibnamefont {Khait}},
  \bibinfo {author} {\bibfnamefont {M.~G.}\ \bibnamefont {Kozlov}}, \bibinfo
  {author} {\bibfnamefont {L.~N.}\ \bibnamefont {Labzowsky}}, \bibinfo {author}
  {\bibfnamefont {A.~O.}\ \bibnamefont {Mitrushenkov}}, \bibinfo {author}
  {\bibfnamefont {A.~V.}\ \bibnamefont {Shtoff}}, \ and\ \bibinfo {author}
  {\bibfnamefont {A.~V.}\ \bibnamefont {Titov}},\ }\href@noop {} {\bibfield
  {journal} {\bibinfo  {journal} {Phys. Lett. A}\ }\textbf {\bibinfo {volume}
  {167}},\ \bibinfo {pages} {280} (\bibinfo {year} {1992})}\BibitemShut
  {NoStop}%
\bibitem [{\citenamefont {Kozlov}\ and\ \citenamefont
  {Ezhov}(1994)}]{kozlov:1994}%
  \BibitemOpen
  \bibfield  {author} {\bibinfo {author} {\bibfnamefont {M.~G.}\ \bibnamefont
  {Kozlov}}\ and\ \bibinfo {author} {\bibfnamefont {V.~F.}\ \bibnamefont
  {Ezhov}},\ }\href@noop {} {\bibfield  {journal} {\bibinfo  {journal} {Phys.
  Rev. A}\ }\textbf {\bibinfo {volume} {49}},\ \bibinfo {pages} {4502}
  (\bibinfo {year} {1994})}\BibitemShut {NoStop}%
\bibitem [{\citenamefont {Kozlov}\ and\ \citenamefont
  {Labzowsky}(1995)}]{kozlov:1995}%
  \BibitemOpen
  \bibfield  {author} {\bibinfo {author} {\bibfnamefont {M.~G.}\ \bibnamefont
  {Kozlov}}\ and\ \bibinfo {author} {\bibfnamefont {L.~N.}\ \bibnamefont
  {Labzowsky}},\ }\href {\doibase 10.1088/0953-4075/28/10/008} {\bibfield
  {journal} {\bibinfo  {journal} {J. Phys. B}\ }\textbf {\bibinfo {volume}
  {28}},\ \bibinfo {pages} {1933} (\bibinfo {year} {1995})}\BibitemShut
  {NoStop}%
\bibitem [{\citenamefont {Kozlov}\ \emph {et~al.}(1997)\citenamefont {Kozlov},
  \citenamefont {Titov}, \citenamefont {Mosyagin},\ and\ \citenamefont
  {Souchko}}]{kozlov:1997}%
  \BibitemOpen
  \bibfield  {author} {\bibinfo {author} {\bibfnamefont {M.~G.}\ \bibnamefont
  {Kozlov}}, \bibinfo {author} {\bibfnamefont {A.~V.}\ \bibnamefont {Titov}},
  \bibinfo {author} {\bibfnamefont {N.~S.}\ \bibnamefont {Mosyagin}}, \ and\
  \bibinfo {author} {\bibfnamefont {P.~V.}\ \bibnamefont {Souchko}},\
  }\href@noop {} {\bibfield  {journal} {\bibinfo  {journal} {Phys. Rev. A}\
  }\textbf {\bibinfo {volume} {56}},\ \bibinfo {pages} {R3326} (\bibinfo {year}
  {1997})}\BibitemShut {NoStop}%
\bibitem [{\citenamefont {Titov}, \citenamefont {Mosyagin},\ and\ \citenamefont
  {Ezhov}(1996)}]{titov:1996}%
  \BibitemOpen
  \bibfield  {author} {\bibinfo {author} {\bibfnamefont {A.~V.}\ \bibnamefont
  {Titov}}, \bibinfo {author} {\bibfnamefont {N.~S.}\ \bibnamefont {Mosyagin}},
  \ and\ \bibinfo {author} {\bibfnamefont {V.~F.}\ \bibnamefont {Ezhov}},\
  }\href@noop {} {\bibfield  {journal} {\bibinfo  {journal} {Phys. Rev. Lett.}\
  }\textbf {\bibinfo {volume} {77}},\ \bibinfo {pages} {5346} (\bibinfo {year}
  {1996})}\BibitemShut {NoStop}%
\bibitem [{\citenamefont {Mosyagin}, \citenamefont {Kozlov},\ and\
  \citenamefont {Titov}(1998)}]{mosyagin:1998}%
  \BibitemOpen
  \bibfield  {author} {\bibinfo {author} {\bibfnamefont {N.~S.}\ \bibnamefont
  {Mosyagin}}, \bibinfo {author} {\bibfnamefont {M.~G.}\ \bibnamefont
  {Kozlov}}, \ and\ \bibinfo {author} {\bibfnamefont {A.~V.}\ \bibnamefont
  {Titov}},\ }\href@noop {} {\bibfield  {journal} {\bibinfo  {journal} {J.
  Phys. B}\ }\textbf {\bibinfo {volume} {31}},\ \bibinfo {pages} {L763}
  (\bibinfo {year} {1998})}\BibitemShut {NoStop}%
\bibitem [{\citenamefont {Quiney}, \citenamefont {Skaane},\ and\ \citenamefont
  {Grant}(1998)}]{quiney:1998}%
  \BibitemOpen
  \bibfield  {author} {\bibinfo {author} {\bibfnamefont {H.~M.}\ \bibnamefont
  {Quiney}}, \bibinfo {author} {\bibfnamefont {H.}~\bibnamefont {Skaane}}, \
  and\ \bibinfo {author} {\bibfnamefont {I.~P.}\ \bibnamefont {Grant}},\
  }\href@noop {} {\bibfield  {journal} {\bibinfo  {journal} {J. Phys. B}\
  }\textbf {\bibinfo {volume} {31}},\ \bibinfo {pages} {L85} (\bibinfo {year}
  {1998})}\BibitemShut {NoStop}%
\bibitem [{\citenamefont {Parpia}(1998)}]{parpia:1998}%
  \BibitemOpen
  \bibfield  {author} {\bibinfo {author} {\bibfnamefont {F.~A.}\ \bibnamefont
  {Parpia}},\ }\href@noop {} {\bibfield  {journal} {\bibinfo  {journal} {J.
  Phys. B}\ }\textbf {\bibinfo {volume} {31}},\ \bibinfo {pages} {1409}
  (\bibinfo {year} {1998})}\BibitemShut {NoStop}%
\bibitem [{\citenamefont {Kozlov}\ and\ \citenamefont
  {Derevianko}(2006)}]{kozlov:2006}%
  \BibitemOpen
  \bibfield  {author} {\bibinfo {author} {\bibfnamefont {M.~G.}\ \bibnamefont
  {Kozlov}}\ and\ \bibinfo {author} {\bibfnamefont {A.}~\bibnamefont
  {Derevianko}},\ }\href@noop {} {\bibfield  {journal} {\bibinfo  {journal}
  {Phys. Rev. Lett.}\ }\textbf {\bibinfo {volume} {97}},\ \bibinfo {pages}
  {63001} (\bibinfo {year} {2006})}\BibitemShut {NoStop}%
\bibitem [{\citenamefont {Nayak}\ and\ \citenamefont
  {Chaudhuri}(2006{\natexlab{a}})}]{nayak:2006}%
  \BibitemOpen
  \bibfield  {author} {\bibinfo {author} {\bibfnamefont {M.~K.}\ \bibnamefont
  {Nayak}}\ and\ \bibinfo {author} {\bibfnamefont {R.~K.}\ \bibnamefont
  {Chaudhuri}},\ }\href {\doibase 10.1016/j.cplett.2005.11.065} {\bibfield
  {journal} {\bibinfo  {journal} {Chem. Phys. Lett.}\ }\textbf {\bibinfo
  {volume} {419}},\ \bibinfo {pages} {191} (\bibinfo {year}
  {2006}{\natexlab{a}})}\BibitemShut {NoStop}%
\bibitem [{\citenamefont {Nayak}\ and\ \citenamefont
  {Chaudhuri}(2006{\natexlab{b}})}]{nayak:2006a}%
  \BibitemOpen
  \bibfield  {author} {\bibinfo {author} {\bibfnamefont {M.~K.}\ \bibnamefont
  {Nayak}}\ and\ \bibinfo {author} {\bibfnamefont {R.~K.}\ \bibnamefont
  {Chaudhuri}},\ }\href {\doibase 10.1088/0953-4075/39/5/020} {\bibfield
  {journal} {\bibinfo  {journal} {J. Phys. B At. Mol. Opt. Phys.}\ }\textbf
  {\bibinfo {volume} {39}},\ \bibinfo {pages} {1231} (\bibinfo {year}
  {2006}{\natexlab{b}})}\BibitemShut {NoStop}%
\bibitem [{\citenamefont {Nayak}\ and\ \citenamefont
  {Chaudhuri}(2007)}]{nayak:2007}%
  \BibitemOpen
  \bibfield  {author} {\bibinfo {author} {\bibfnamefont {M.~K.}\ \bibnamefont
  {Nayak}}\ and\ \bibinfo {author} {\bibfnamefont {R.~K.}\ \bibnamefont
  {Chaudhuri}},\ }\href {\doibase 10.1088/1742-6596/80/1/012051} {\bibfield
  {journal} {\bibinfo  {journal} {J. Phys. Conf. Ser.}\ }\textbf {\bibinfo
  {volume} {80}},\ \bibinfo {pages} {012051} (\bibinfo {year}
  {2007})}\BibitemShut {NoStop}%
\bibitem [{\citenamefont {Nayak}\ and\ \citenamefont
  {Chaudhuri}(2009)}]{nayak:2009a}%
  \BibitemOpen
  \bibfield  {author} {\bibinfo {author} {\bibfnamefont {M.~K.}\ \bibnamefont
  {Nayak}}\ and\ \bibinfo {author} {\bibfnamefont {R.~K.}\ \bibnamefont
  {Chaudhuri}},\ }\href {\doibase 10.1007/s12043-009-0110-z} {\bibfield
  {journal} {\bibinfo  {journal} {Pramana - J. Phys.}\ }\textbf {\bibinfo
  {volume} {73}},\ \bibinfo {pages} {581} (\bibinfo {year} {2009})}\BibitemShut
  {NoStop}%
\bibitem [{\citenamefont {Abe}\ \emph {et~al.}(2014)\citenamefont {Abe},
  \citenamefont {Gopakumar}, \citenamefont {Hada}, \citenamefont {Das},
  \citenamefont {Tatewaki},\ and\ \citenamefont {Mukherjee}}]{abe:2014}%
  \BibitemOpen
  \bibfield  {author} {\bibinfo {author} {\bibfnamefont {M.}~\bibnamefont
  {Abe}}, \bibinfo {author} {\bibfnamefont {G.}~\bibnamefont {Gopakumar}},
  \bibinfo {author} {\bibfnamefont {M.}~\bibnamefont {Hada}}, \bibinfo {author}
  {\bibfnamefont {B.~P.}\ \bibnamefont {Das}}, \bibinfo {author} {\bibfnamefont
  {H.}~\bibnamefont {Tatewaki}}, \ and\ \bibinfo {author} {\bibfnamefont
  {D.}~\bibnamefont {Mukherjee}},\ }\href {\doibase 10.1103/PhysRevA.90.022501}
  {\bibfield  {journal} {\bibinfo  {journal} {Phys. Rev. A - At. Mol. Opt.
  Phys.}\ }\textbf {\bibinfo {volume} {90}},\ \bibinfo {pages} {1} (\bibinfo
  {year} {2014})},\ \Eprint {http://arxiv.org/abs/1405.0544} {arXiv:1405.0544}
  \BibitemShut {NoStop}%
\bibitem [{\citenamefont {Meyer}, \citenamefont {Bohn},\ and\ \citenamefont
  {Deskevich}(2006)}]{meyer:2006}%
  \BibitemOpen
  \bibfield  {author} {\bibinfo {author} {\bibfnamefont {E.~R.}\ \bibnamefont
  {Meyer}}, \bibinfo {author} {\bibfnamefont {J.~L.}\ \bibnamefont {Bohn}}, \
  and\ \bibinfo {author} {\bibfnamefont {M.~P.}\ \bibnamefont {Deskevich}},\
  }\href {\doibase 10.1103/PhysRevA.73.062108} {\bibfield  {journal} {\bibinfo
  {journal} {Phys. Rev. A - At. Mol. Opt. Phys.}\ }\textbf {\bibinfo {volume}
  {73}},\ \bibinfo {pages} {1} (\bibinfo {year} {2006})},\ \Eprint
  {http://arxiv.org/abs/0604205} {arXiv:0604205 [physics]} \BibitemShut
  {NoStop}%
\bibitem [{\citenamefont {Knight}(1971)}]{knight:1971}%
  \BibitemOpen
  \bibfield  {author} {\bibinfo {author} {\bibfnamefont {L.~B.}\ \bibnamefont
  {Knight}},\ }\href {\doibase 10.1063/1.1674610} {\bibfield  {journal}
  {\bibinfo  {journal} {J. Chem. Phys.}\ }\textbf {\bibinfo {volume} {54}},\
  \bibinfo {pages} {322} (\bibinfo {year} {1971})}\BibitemShut {NoStop}%
\bibitem [{\citenamefont {Ryzlewicz}\ \emph {et~al.}(1982)\citenamefont
  {Ryzlewicz}, \citenamefont {Sch{\"{u}}tze-Pahlmann}, \citenamefont {Hoeft},\
  and\ \citenamefont {T{\"{o}}rring}}]{ryzlewicz:1982}%
  \BibitemOpen
  \bibfield  {author} {\bibinfo {author} {\bibfnamefont {C.}~\bibnamefont
  {Ryzlewicz}}, \bibinfo {author} {\bibfnamefont {H.~U.}\ \bibnamefont
  {Sch{\"{u}}tze-Pahlmann}}, \bibinfo {author} {\bibfnamefont {J.}~\bibnamefont
  {Hoeft}}, \ and\ \bibinfo {author} {\bibfnamefont {T.}~\bibnamefont
  {T{\"{o}}rring}},\ }\href {\doibase 10.1016/0301-0104(82)85045-3} {\bibfield
  {journal} {\bibinfo  {journal} {Chem. Phys.}\ }\textbf {\bibinfo {volume}
  {71}},\ \bibinfo {pages} {389} (\bibinfo {year} {1982})}\BibitemShut
  {NoStop}%
\bibitem [{\citenamefont {{Van Zee}}\ \emph {et~al.}(1978)\citenamefont {{Van
  Zee}}, \citenamefont {Seely}, \citenamefont {DeVore},\ and\ \citenamefont
  {{Weltner Jr.}}}]{vanzee:1978}%
  \BibitemOpen
  \bibfield  {author} {\bibinfo {author} {\bibfnamefont {R.~J.}\ \bibnamefont
  {{Van Zee}}}, \bibinfo {author} {\bibfnamefont {M.~L.}\ \bibnamefont
  {Seely}}, \bibinfo {author} {\bibfnamefont {T.~C.}\ \bibnamefont {DeVore}}, \
  and\ \bibinfo {author} {\bibfnamefont {W.}~\bibnamefont {{Weltner Jr.}}},\
  }\href@noop {} {\bibfield  {journal} {\bibinfo  {journal} {J. Phys. Chem.}\
  }\textbf {\bibinfo {volume} {82}},\ \bibinfo {pages} {1978} (\bibinfo {year}
  {1978})}\BibitemShut {NoStop}%
\bibitem [{\citenamefont {Nayak}\ and\ \citenamefont {Das}(2009)}]{nayak:2009}%
  \BibitemOpen
  \bibfield  {author} {\bibinfo {author} {\bibfnamefont {M.~K.}\ \bibnamefont
  {Nayak}}\ and\ \bibinfo {author} {\bibfnamefont {B.~P.}\ \bibnamefont
  {Das}},\ }\href {\doibase 10.1103/PhysRevA.79.060502} {\bibfield  {journal}
  {\bibinfo  {journal} {Phys. Rev. A}\ }\textbf {\bibinfo {volume} {79}},\
  \bibinfo {pages} {060502} (\bibinfo {year} {2009})}\BibitemShut {NoStop}%
\bibitem [{\citenamefont {Knight}(1981)}]{knight:1981}%
  \BibitemOpen
  \bibfield  {author} {\bibinfo {author} {\bibfnamefont {L.~B.}\ \bibnamefont
  {Knight}},\ }\href {\doibase 10.1063/1.441040} {\bibfield  {journal}
  {\bibinfo  {journal} {J. Chem. Phys.}\ }\textbf {\bibinfo {volume} {74}},\
  \bibinfo {pages} {6009} (\bibinfo {year} {1981})}\BibitemShut {NoStop}%
\end{thebibliography}
%
\end{document}